\documentclass[a4paper,fleqn,review]{cas-sc}

\usepackage{cleveref}
\usepackage[normalem]{ulem}

\usepackage[numbers,sort&compress]{natbib}%\usepackage[authoryear]{natbib}

\usepackage{xcolor}

\usepackage{bm} % boldface for greek symbols

\def\tsc#1{\csdef{#1}{\textsc{\lowercase{#1}}\xspace}}
\tsc{WGM}
\tsc{QE}
\newdefinition{rmk}{Remark}
\newproof{pf}{Proof}
\newproof{pot}{Proof of Theorem \ref{thm}}
\begin{document}

\let\WriteBookmarks\relax
\def\floatpagepagefraction{1}
\def\textpagefraction{.001}

\shorttitle{An interpretable  wildfire spreading model for real-time predictions}    

\shortauthors{Vogiatzoglou, et al.,}    

\title [mode = title]{An interpretable  wildfire spreading model for real-time predictions}  

%\tnotemark[1,2] 

%  \tnotetext[1]{This document is the results of the research project
%    funded by the National Science Foundation.}

%  \tnotetext[2]{The second title footnote which is a longer text
%    matter to fill through the whole text width and overflow into
%    another line in the footnotes area of the first page.}

\author[1]{K. Vogiatzoglou}%[options]

%\fnmark[footnote mark no]

\ead{kvogiatzoglou@uth.gr}

%\ead[url]{<URL>}

\credit{Conceptualization, Software, Validation, Formal analysis, Data Curation, Writing - Review \& Editing, Visualization}

\affiliation[1]{organization={Department of Mechanical Engineering, University of Thessaly},
%            addressline={Pedio}, 
            city={Volos},
%          citysep={}, % Uncomment http://www.ebay.com/if no comma needed between city and postcode
            postcode={38334}, 
%            state={xxx},
            country={Greece}
}

\affiliation[2]{organization={Computational Science and Engineering Laboratory, School of Engineering and Applied Science, Harvard University},
%            addressline={29 Oxford Street}, 
            city={Cambridge},
%          citysep={}, % Uncomment if no comma needed between city and postcode
            postcode={MA 02138}, 
%            state={xxx},
            country={USA}
}

\author[1]{C. Papadimitriou}%[<options>]

% Footnote of the second author

% Email id of the second author
\ead{costasp@uth.gr}

%% URL of the second author

%% Credit authorship
%  \credit{Conceptualization of this study, Methodology}

%\affiliation[aff no]{organization={},
%%          citysep={}, % Uncomment if no comma needed between city and postcode
%            country={}
%}

%% Corresponding author text
%\cortext[1]{Corresponding author}
%
%% Footnote text


\author[1]{K. Ampountolas}

% Footnote of the second author

% Email id of the second author
\ead{k.ampountolas@uth.gr}

%% URL of the second author

%% Credit authorship
%  \credit{Conceptualization of this study, Methodology}

\author[2]{M. Chatzimanolakis}%[<options>]

% Footnote of the second author

% Email id of the second author
\ead{mchatzimanolakis@seas.harvard.edu}

\author[2]{P. Koumoutsakos}%[<options>]

% Footnote of the second author

% Email id of the second author
\ead{petros@seas.harvard.edu}

\author[1]{V. Bontozoglou}%[<options>]
[orcid=0000-0002-3342-0293]

%% Corresponding author text
\cortext[1]{Corresponding author}

\cormark[1]

% Footnote of the second author

% Email id of the second author
\ead{bont@uth.gr}

%% URL of the second author

%% Credit authorship
%  \credit{Conceptualization of this study, Methodology}

\begin{abstract}
Forest fires are a key component of natural ecosystems, but their increased frequency and intensity have devastating social, economic, and environmental implications. Thus, there is a great need for trustworthy digital tools capable of providing real-time estimates of fire evolution and human interventions. This work develops an interpretable, physics-based model that will serve as the core of a broader wildfire prediction tool. The modeling approach involves a simplified description of combustion kinetics and thermal energy transfer (averaged over local plantation height) and leads to a computationally inexpensive system of differential equations that provides the spatiotemporal evolution of the two-dimensional fields of temperature and combustibles. Key aspects of the model include the estimation of mean wind velocity through the plantation and the inclusion of the effect of ground inclination. Predictions are successfully compared to benchmark literature results concerning the effect of flammable bulk density, moisture content, and the combined influence of wind and slope. Simulations appear to provide qualitatively correct descriptions of firefront propagation from a localized ignition site in a homogeneous or heterogeneous canopy, of acceleration resulting from the collision of oblique firelines, and of firefront overshoot or arrest at fuel break zones. 
\end{abstract}


\begin{highlights}
\item New physics-based model with simplified reaction kinetics and energy balance. 

\item Modest parameter set captures key physical quantities and heat transport mechanisms.

\item Model validated by predicting the effect of fuel properties, wind and inclination.

\item Simulations of firefront collision and interaction with fuel variations and breaks.
\end{highlights}

% Keywords

\maketitle

% Main text

\section{Introduction}
\label{sec:Introduction}

In recent years, the frequency and severity of natural disasters resulting from wildfires have witnessed an alarming rise. This trend is expected to persist in the future, primarily due to the combined effects of climate change and unfavorable human activities \citep{MDFlannigan2000,ACarvalho2011}. Wildfires, which are rapidly spreading fires engulfing expansive vegetated areas, pose a major threat to communities at the wildland-urban interface \citep{Juliano2021}. The primary concern revolves around the threat to human life, but the detrimental consequences of this natural hazard are evident across social, economic, and environmental domains \citep{NegarElhamiKhorasani2022}. The most crucial information needed to safeguard lives and property and guide aggressive firefighter suppression efforts is the rate of spread ($\rm ROS$) of the firefront. 
Moreover, the detection of the fireline perimeter, along with the evaluation of the fire power and intensity, represent essential quantities to anticipate during events of excessive fire propagation \citep{AlexanderME1985}.

Wildland fires are intricate environmental phenomena that integrate multiple spatiotemporal scales and physical processes, including the chemistry of fuel combustion, the physics of fluid flow and heat-mass transport, while being subjected to dynamic atmospheric conditions \citep{Sullivan2009, Goodrick2022}. In forests and shrublands, the fuel consists of particle-like materials (e.g., leaves, grass, twigs, and pine needles) of varying size and composition. The heterogeneity of the biomass fuel in a forest area can be considered a random field, contributing to the uncertain flame direction. The heating of the fuel particles, which initiates combustion, is evidently provided by radiation and convection, though their relative contributions are under debate \citep{Finney2013, FrankmanD2013}. Heat convection is influenced by airflow above the plant canopy, dictated by local meteorological conditions and topographical features \citep{Finnigan2007, Finnigan2010}, and further affected by convection currents triggered by flame instabilities \citep{Finney2015}.

Over the years, numerous modeling approaches have been proposed to develop hypotheses on how fires grow and spread, yet many questions remain regarding the underlying physics of these phenomena \citep{JoaoSilva2022}. Models predicting wildland fire dynamics, particularly the $\rm ROS$, may be classified into three categories. At the simplest and most operation-oriented end are semi-empirical models that establish functional relations between the $\rm ROS$ and key parameters such as wind speed, slope, fuel bulk density, size, and moisture content \citep{Weise1997, Vega1998, Fernandes2001, Marino2012, Cruz2020, EPastor2003}.
At the other end of the spectrum are three-dimensional (3D) computational fluid dynamics (CFD) simulators aiming to comprehensively describe all physical and chemical interactions over a broad range of scales. Examples include multiphase models \citep{Goodrick2022, Morvan2004} that analyze combustion processes at scales of tens of centimeters, wildfire propagation models \citep{Linn2005, JMCanfield2014} like FIRETEC and FlamMap at meter scales, and atmospheric boundary layer models \citep{Mandel2011, Coen2013} spanning hundreds of meters to provide an overall view of fire progression. The CFD models are computationally intensive, so their main contribution at present is the elucidation of the governing physical mechanisms operative at the various spatiotemporal scales.

Between empirical functional relations and CFD tools, several "simple but interpretable" models have been suggested to describe the fundamental physics while remaining computationally tractable, with the ultimate objective of providing real-time predictive capabilities \citep{Coen2007, Mandel2008, Simeoni2011, Buerger2020}. The present work develops such a model, using simplified reaction kinetics and energy balances averaged over the plantation height. The model is based on a small number of parameters that represent crucial physical quantities and heat transport mechanisms, and is presently validated by comparison to benchmark literature results.

It is envisioned that this model may be used in the future as the core of a risk-informed decision tool for ongoing fires, which will be continuously fed with remote sensing data \citep{Nicholas2019, Lareau2022, Paugam2012} and will provide updated optimal estimates of the model parameters. Such an approach is expected to cope successfully with the stochastic and epistemic uncertainties arising from modeling assumptions and environmental variations.

The remainder of the paper is structured as follows. \Cref{sec:Modeling} provides a concise overview of the wildfire model derivation and relevant computations. In \Cref{sec:Validation}, the model is validated with literature results for one-dimensional fire propagation (infinitely long, straight frontline) and is subsequently used to investigate a variety of two-dimensional propagation case studies. Finally, conclusions are drawn, and future refinements are outlined in \Cref{sec:Conclusions}.

\section{Modeling}
\label{sec:Modeling}

The proposed model incorporates several primary physical mechanisms that influence the rate of spread of a firefront. (i) Reaction kinetics for water evaporation and wood combustion are described by first-order Arrhenius rates, the latter empirically adjusted for the limited availability of oxygen. In the current stage of development, the kinetics are confined to a single exemplary fuel. However, the formulation can be readily extended to include an arbitrary number of fuel components, as described by the  available standard fuel mixtures \citep{Anderson1982, Scott2005}. (ii) Heat transfer by convection, due to the motion of the gaseous phase (air/flue gases) through the plantation, is estimated using a simple fluid dynamic argument based on the dominant role of drag provided by a relatively dense canopy. This approach, which can be substituted by more elaborate models in the future, predicts the mean velocity through the plantation, incorporating as input the upstream wind speed 10 m above the ground. (iii) Localized heat transfer around the flame by turbulent transport and radiation is simulated by an effective dispersion coefficient, which increases linearly with the mean velocity through the canopy. Finally, heat losses to the surroundings are estimated by combining free convection and radiation from a horizontal surface.

\subsection{Problem setup and reaction kinetics}
\label{sec:Mass_Reaction}

A field extending in the streamwise ($x-$) and lateral ($y-$) direction is considered, covered by a plantation of uniform height $H$ ([=] m). The bulk density of solid material, $m_{s,0}=\alpha\rho_s$ ([=] kg/m$^3$), is defined as the product of the packing ratio of solids, $\alpha$ ([=] solid m$^3$/total m$^3$), and the material density $\rho_s$ ([=] kg/m$^3$). The solid phase comprises two components: water ($m_{s1}$) and combustibles ($m_{s2}$), with the latter including non-aqueous volatiles (e.g., CO, CO$_2$, NO$_x$, VOC) and charcoal. Thus, $m_{s,0}=m_{s1,0}+m_{s2,0}$, where $m_{s1,0}/m_{s,0}$ represents the original water fraction in the plantation and $m_{s2,0}/m_{s,0}$ denotes the initial fraction of combustibles. Equivalently, the fuel moisture content (FMC), traditionally defined on a dry basis, is ${\rm FMC}=100\, (m_{s1,0}/m_{s2,0})$. Water content is typically classified as the humidity of live and dead plants. While the former varies mainly with plant type and season of the year, the latter is a function of air humidity and fuel size (i.e., the smaller the fuel particle, the faster it equilibrates with air humidity). With decreasing moisture content, the fuel is more flammable, thereby accelerating fire propagation \citep{Rossa2017, KhalidMoinuddin2021}.

During the combustion process, the remaining mass of solid material is $m_s(x,y,t)=m_{s1}(x,y,t)+m_{s2}(x,y,t)$. The rest of the field volume is occupied by the gaseous phase, whose mass, $m_g$ ([=] kg/m$^3$), increases over time as solid material converts into gaseous products. Therefore, the conservation of total mass for a closed system is expressed as:
\begin{equation}
\label{eq:MassBalance}
m_s+m_g=\alpha\rho_s+(1-\alpha)\rho_g,
\end{equation}
where $\rho_g$ ([=] kg/m$^3$) is a representative gas density. \Cref{eq:MassBalance} may be written in dimensionless form by defining the remaining fuel mass fraction of water, $S_1=m_{s1}/m_{s,0}$ (endothermic phase), and combustibles, $S_2=m_{s2}/m_{s,0}$ (exothermic phase). The sum of these fractions, $S=S_1+S_2$, represents the total fuel mass fraction at any given instant. Additionally, $S_g=m_g/m_{s,0}$ denotes the dimensionless gaseous phase mass fraction. Thus, \cref{eq:MassBalance} in dimensionless form is:
\begin{equation}
\label{eq:MassBalanceDimless}
S+S_g=1+\frac{1-\alpha}{\alpha}\lambda,
\end{equation}
with $\lambda$ representing the density ratio $\rho_g/\rho_s$. It is noteworthy that parameter $\alpha$ is a small number, typically on the order of $[10^{-3},10^{-2}]$ \citep{Pimont2009}.

As a simplified overview of a very complex set of chemical and physical processes \citep{Sullivan2017}, combustion may be considered to initiate with fuel dehydration, followed by pyrolysis and charring reactions, and concluding with char oxidation. While dehydration (T $<$ 500 K) is endothermic, the start of pyrolysis reactions (T $>$ 550-650 K) marks the onset of the flammable exothermic part \citep{FranciscoJSeron2005, RubenSudhakarDhanarathinam2011}. Oxygen delivery rate, influenced by turbulence and buoyant instabilities, becomes critical during char oxidation, resulting in flaming or glowing-smoldering combustion \citep{SantosoMA2019}.

Wood dehydration, and disintegration and combustion are modeled by two consecutive reactions following first-order Arrhenius kinetics. Thus, the variation in water content with time is expressed in the following form:
\begin{equation}
\label{eq:Reaction1}
\frac{\partial S_1}{\partial t} =-S_1\,r_1.
\end{equation}
The reaction rate for the endothermic phase
is modeled as follows: 
\begin{equation}
\label{eq:Arrhenius1}
r_1 =c_{s1}\,e^{-\frac{b_1}{T}},
\end{equation}
with $T$ ([=] K) representing the temperature of the fire layer and parameters $c_{s1}$ ([=] s$^{-1}$) and $b_1$ ([=] K) quantifying differences in behavior between dead and live moisture content \citep{Rossa2017}. A similar expression with different constants is used for the rate $r_2$ of the burning process. However, in this case, the reaction rate is limited not only by low temperature but also by a lack of oxygen \citep{Leckner1999}. Considering these two resistances in series, the final reaction rate is modeled as:
\begin{equation}
\label{eq:Arrhenius2}
r_{2t} = \frac{r_2\,r_m}{r_2 + r_m},
\end{equation}
where $r_m$ ([=] s$^{-1}$) is the rate of oxygen arrival to the burning solid. Thus, the temporal variation of combustibles is given by:
\begin{equation}
\label{eq:Reaction2}
\frac{\partial S_2}{\partial t} =-S_2\,r_{2t} =-S_2\,\frac{c_{s2}\,e^{-\frac{b_2}{T}}r_m}{c_{s2}\,e^{-\frac{b_2}{T}}+r_m}.
\end{equation}
Parameters $c_{s2}$ ([=] s$^{-1}$) and $b_2$ ([=] K) depend on fuel characteristics and determine the intensity of pyrolysis and combustion. The rate of oxygen delivery, $r_m$, is empirically determined to take values in the range $[10^{-3},10^{-2}]$ since very small values result in fire extinction, while very large values result in unrealistically high flame temperatures. It is an increasing function of the mean gaseous velocity through the canopy, $\left<u\right>$ ([=] m/s) (to be calculated in \Cref{sec:WindSpeed}), and is presently approximated as $r_m = r_{m,0} + r_{m,c}(\left<{u}\right> - 1)$. 

\subsection{Thermal energy balance}
\label{sec:EnergyBalance}

The energy balance is based on the assumption of local equilibrium \citep{Simeoni2001}, i.e., that the solid and gaseous phases have the same temperature. As a result, we may write the following equation for a volume of unit surface area and height $H$:
\begin{equation}
\label{eq:EnergyBalance}
(m_s\, c_{ps}+m_g\, c_{pg})\frac{\partial T}{\partial t}=
-A_1\,m_{s1}\,r_1+A_2\,m_{s2}\,r_{2t}-m_g c_{pg}\left<\mathbf{u}\right>\cdot \mathbf{\nabla}T+
m_g c_{pg}\nabla\cdot\left(\mathbf{D_{\rm eff}}\cdot\nabla T\right)-\frac{U}{H}(T-T_a).
\end{equation}
The coefficients $c_{ps}$ and $c_{pg}$ ([=] J/kgK) are the heat capacities of the solid and gaseous phases, while $A_1$ and $A_2$ ([=] J/kg) are the standard heats of the endothermic (water evaporation) and exothermic (pyrolysis and combustion) reactions. The term $\left<\mathbf{u}\right>$ is the mean velocity vector through the plantation, and the respective term is the convective contribution to the energy balance. The term $\mathbf{D_{\rm eff}}$ ([=] m$^2$/s) is a dispersion coefficient vector, and $U$ ([=] W/m$^2$K) is an overall heat transfer coefficient for thermal losses to the environment. Here, $T_a$ represents the ambient temperature, and $\nabla$ denotes the gradient operator.

It is noted at the onset that the velocity $\left<\mathbf{u}\right>=(u_x,u_y)$, which represents gas flow averaged over the plantation height, is chosen by ad-hoc arguments related to the resistance to flow exerted by the canopy and thus may vary according to the spatial variation of canopy density. Most importantly, it need not satisfy the continuity equation, as our model is two-dimensional and does not explicitly include motions in the vertical direction. Indeed, spatial variations of $\left<\mathbf{u}\right>$ through the firefront imply the existence of a vertical component, e.g., the buoyant plume. This treatment decouples the energy from the fluid mechanics aspect of the problem and contributes to the computational efficiency of our model. As a next step in future work, improved choices of the mean velocity field through the canopy may be provided by an independent study of the fluid mechanics.

The dispersion coefficient quantifies short-range heat transport by radiation, buoyant instabilities, and turbulent eddies. Radiation has long been considered the main transport mechanism \citep{Sullivan2009, Simeoni2011} for preheating to ignite the unburned fuel in front of the flame. However, this view was questioned \citep{Finney2013}, and it has been argued that instabilities caused by buoyant dynamics and unsteady convection (flame intermittency) dominate local transport \citep{Finney2015} when the wind is high enough for the flame to be tilted.

According to dimensional analysis \citep{Cussler}, the dispersion coefficient is proportional to the product of the local velocity and length scales. In the present case, the characteristic scales may be different in the $x-$ and $y-$directions due to the effect of wind, and therefore the dispersion coefficient is taken as a 2x2 diagonal matrix with components $D_{\rm eff,x}$ and $D_{\rm eff,y}$ computed as follows: 
\begin{equation}
\label{eq:Dispersion}
D_{\rm eff,x}=D_{\rm rb} + A_d\left<u_{x}\right> L_{\rm x}\left(1-e^{-\gamma_d w_{\rm x}}\right) , \quad D_{\rm eff,y}=D_{\rm rb} + A_d\left<u_y\right> L_{\rm y}\left(1-e^{-\gamma_d w_{\rm y}}\right),
\end{equation}
The first term, $D_{\rm rb}$ ([=] m$^2$/s), in \cref{eq:Dispersion} is the contribution of radiation and buoyancy in the absence of wind, and the second is the intensification of dispersion due to the wind. Two new length vectors, $\mathbf{L}$ and $\mathbf{w}$, are introduced in \cref{eq:Dispersion}, which represent global lengthscales of the burning field and are updated at every time step of the computation. The vector $\mathbf{L}=(L_{\rm x},\,L_{\rm y})$ is the distance from the location of global maximum temperature over the field, $T_{\rm max}$, to the points in the $x-$ and $y-$direction where temperature has dropped down to $T=0.1\,T_{max} + T_a$. Thus, it provides a representative measure of the width of the burning zone in each direction.

The term in parenthesis on the right-hand side of \cref{eq:Dispersion} expresses a mitigation of the effect of wind on the dispersion coefficient when the fire is spatially restricted, i.e., when the fireline is short \citep{CheneyNP1995}. It is known that the buoyant column of rising gases partially obstructs the ambient air flow and, as a result, pressure gradients develop that tend to redirect the flow around the fireline. However, as has been shown by numerical simulations \citep{JMCanfield2014}, a longer fireline is not easily bypassed, and thus the air is forced to go through the plume.

The fireline length is computed at every time step as the length of the vector $\mathbf{w}=(w_{\rm x},\,w_{\rm y})$, defined as follows: the temperature maxima in one direction (say $x-$) are identified, and their variation in the other direction (say $y-$) is considered. The resulting array, $T_{\rm max}(x_i,y_i)$, is used to locate the endpoints $(x_A,y_A)$ and $(x_B,y_B)$ by the condition $T_{\rm max}>550\,K$. These endpoints define the vector $\mathbf{w}$, and thus $w_{\rm x}=|x_A-x_B|,\,w_{\rm y}=|y_A-y_B|$. Varying the parameter $\gamma_d$ ([=] m$^{-1}$) in \cref{eq:Dispersion} increases or decreases the fireline length at which the asymptotic limit of the $\rm ROS$ of a long fireline is practically reached.

Term $U$ ([=] W/m$^2$K) is an overall heat transfer coefficient that includes free convection and radiation, according to the expression:
\begin{equation}
\label{eq:TotalCoefficient}
U=h_{nc}+\varepsilon\sigma_{b} \left(T^2+T_a^2\right)(T+T_a)=A_{nc}(T-T_a)^{1/3}+\varepsilon\sigma_{b} \left(T^2+T_a^2\right)(T+T_a),
\end{equation}
where $\varepsilon$ is the surface emissivity, $\sigma_b$ is the Stefan-Boltzmann constant, and  $A_{nc}$ ([=] W/m$^2$K$^{4/3})$ sums up terms from the correlation $Nu_{nc}=0.15\,Gr^{1/3}Pr^{1/3}$, valid for a horizontal hot surface and $Gr>10^7$. Invoking the definitions of Nusselt and Grashof numbers:
\begin{equation}
\label{eq:NaturalConvection}
h_{nc} =\underbrace{\left[0.15\left(\frac{g\beta}{\nu^2}\right)^{1/3}\,Pr^{1/3}k \right]}_{\text{$A_{nc}$}}(T-T_a)^{1/3}.  
\end{equation}

\subsection{Gas flow through the canopy}
\label{sec:WindSpeed}

The transfer processes involved in the energy balance depend strongly on the motion of the gaseous phase over and through the canopy, which is expressed as a mean vector velocity, $\left<u\right>$, across the plantation height, $H$. This velocity is determined by the wind speed above the plantation and the resistance to air motion exerted by the plantation. More specifically, \citep{Inoue1963, Finnigan2007} the velocity field of the atmospheric boundary layer exerts a shear, $\tau$ ([=] N/m$^2$), at the top of the canopy and imparts momentum to the underlying air. However, this momentum dissipates not only on the ground but throughout the canopy. In the following, the wind intensity is defined by the velocity, $u_{10}$, at $10$ m above the ground, as it is standard for meteorological measurements.

\begin{figure}[tbp]
\centering
\includegraphics[width=.55\textwidth]{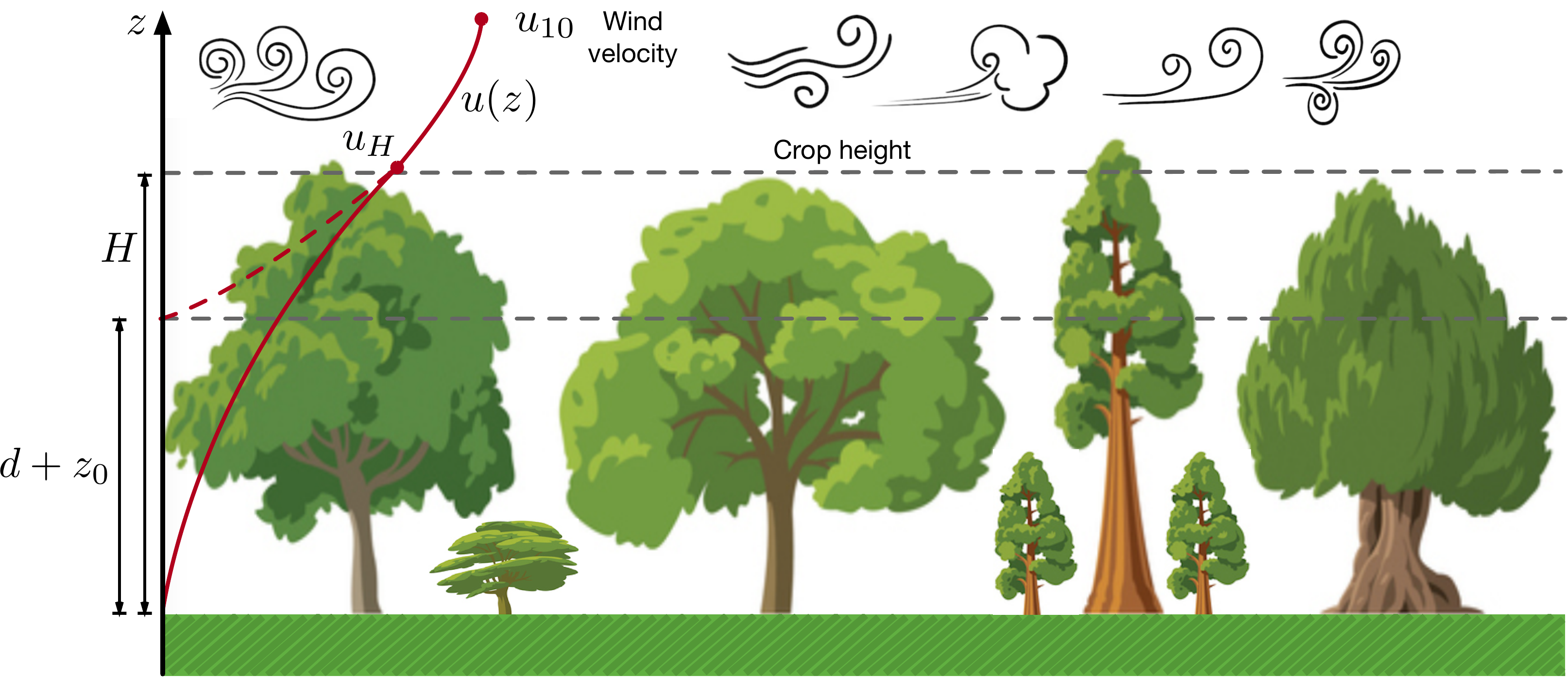}
\caption{Wind field distribution over and through the canopy.}
\label{fig:Plantation}
\end{figure}

Following Inoue \citep{Inoue1963} (see \cref{fig:Plantation}), the turbulent flow above the plantation is described by the following expression (law of the wall), and the nominal wind speed, $u_{10}$, is recovered by setting the height $z=10$ m:
\begin{equation}
\label{eq:LawWall}
u_v(z)=\frac{u_{v*}}{\kappa}\,\ln\left(\frac{z-d}{z_0}\right).
\end{equation}
In \cref{eq:LawWall}, $\kappa=0.41$ is the Karman's constant, $z_0$ ([=] m) is the surface roughness at the top of the plantation \citep{Stull} and $u_{v*}$ ([=] m/s) is the friction velocity, which is evaluated by the substitution $u_v(z=$10 m$)=u_{10}$. The height $d$ ([=] m) corresponds to the "nominal ground" as experienced empirically by the air flow above the plantation. In other words, the tentative extension of the logarithmic velocity profile inside the plantation goes to zero at $z=d+z_0$. This is also the location where the "nominal wall shear stress" $\tau_w$, applies. The height $d$ decreases with wind speed, i.e., with increasing $u_{10}$ the effect of the wind penetrates deeper inside the plantation. This tendency is currently approximated as follows:
\begin{equation}
\label{eq:GroundDepth}
 d=(H-z_0)-\delta\, u_{10},
\end{equation}
where $z_0$ ([=] m) and $\delta$ ([=] s) are expected to vary with the thickness of the plantation (see \cref{table:WinsSpeedParameters} for suggested values).

\begin{table}[htbp]
\centering
\caption{Representative values of the parameters $z_0$, and $\delta$ for a sparse and a dense type of canopy.}
\label{table:WinsSpeedParameters}
\begin{tabular}{ccc}
\toprule
\textbf{Canopy} & $z_{0}$ ([=] m) & $\delta$ ([=] s) \\
\midrule
\textbf{Sparse} & 0.5 & 0.08 \\
\textbf{Dense} & 0.25 & 0.04 \\
\bottomrule
\end{tabular}
\end{table}

The actual velocity profile inside the plantation is determined by the following force balance over a differential section:
\begin{equation}
\label{eq:DragForce}
\frac{d\tau}{dz}=\rho_g C_d A_{pl}u^2,
\end{equation}
where $C_d$ is the drag coefficient \citep{Gonçalves2023}, and $A_{pl}$ ([=] m$^{-1}$) is the total surface area that exerts drag on the flow, per unit control volume ($A_{pl}$ is usually expressed as the product $A_{pl}=\alpha\,s_{pl}$, where $s_{pl}$ is the area per solid volume ratio \citep{Pimont2012}). Invoking the mixing-length hypothesis as a simple turbulence closure model, we may write:
\begin{equation}
\label{eq:MixingLength}\tau=l_m^2\,\left(\frac{du}{dz}\right)^2,
\end{equation}
where it has been argued \citep{Inoue1963, Finnigan2007} that the fluid dynamics inside the plantation are satisfactorily captured by assuming a constant eddy size, $l_m$ ([=] m). Combining \cref{eq:DragForce} with \cref{eq:MixingLength} and setting $u(H)=u_H$ for the velocity at the top of the plantation leads to the following exponential profile inside the canopy:
\begin{equation}
\label{eq:PlantProfile}
u_v(z)=u_H e^{-\eta(1-z/H)},
\end{equation}
where,
\begin{equation}
\label{eq:Eta}
\eta=H\left(\frac{C_d A_{pl}}{2\,l_m^2}\right)^{1/3}, 
\end{equation}
with typical values in the range $\eta\in [2, 3]$, see  \citep{Inoue1963}. Finally, the mean gaseous velocity across the plantation height is calculated as:
\begin{equation}
\label{eq:MeanVelocity}  \left<u_v\right>=\frac{1}{H}\int_0^H u_v(z)dz=\frac{u_H}{\eta}\left(1-e^{-\eta}\right).
\end{equation}

The mean velocity $\left<u_v\right>$ calculated above describes the air motion ahead of the firefront and into the unburned region and is thus relevant to the rate of fire spread. However, behind the front, the air flow meets less drag resistance, as foliage and branches have been to a large extent eliminated by the fire. This latter air flow must thus be described by a higher mean velocity value, $\left<u_b\right>$. Of course, the streamwise gradient of velocity thus imposed will generate pressure gradients that will drive the excess air flow around and over the fireline \citep{JMCanfield2014}.

In order to express the aforementioned effect, we consider the logarithmic velocity profile over bare ground of roughness $z_0$:  
\begin{equation}
\label{eq:LawWallBurning}
u_b(z)=\frac{u_{b*}}{\kappa}\,\ln\left(\frac{z}{z_0}\right),
\end{equation}
and define the mean velocity, $\left<u_b\right>$, over the height $H$ that corresponds to the case of a burned plantation behind the firefront. Thus: 
\begin{equation}
\label{eq:MeanVelocityBurning} \left<u_b\right>=\frac{1}{H-z_0}\int_{z_0}^H u_b(z)dz=\frac{u_{b*}}{\kappa}\left[\frac{H}{H-z_0}\ln\left(\frac{H}{z_0}\right) - 1\right],
\end{equation}
where the friction velocity, $u_{b*}$, is evaluated by the substitution $u_b(z=10 m)=u_{10}$. As $\left<u_v\right>$ is relevant to the intact and $\left<u_b\right>$ to the totally burned-down canopy, we define the varying streamwise air velocity as a function of the instantaneous remaining fraction of combustibles, $x_c=S_2/S_{2,0}$, with $S_{2,0}$ representing the initial composition of the exothermic fuel mass fraction. In particular, we presently use a simple linear approximation, and thus the mean local velocity to be applied in \cref{eq:EnergyBalance} is calculated as:
\begin{equation}\label{eq:MeanTotalVelocity} \left<u\right>=\left<{u_v}\right> + \left(\left<{u_b}\right>-\left<{u_v}\right>\right)(1-x_c).
\end{equation}

More refined estimates of the variation of mean velocity through the firefront may result from consideration of the effect of the dimensionless convection (Byram) number, $N_c$, which compares the free-stream wind speed to the velocity imposed by buoyancy \citep{Nelson2015}.

\subsection{Fire spread on inclined terrain}
\label{sec:Inclination}

Terrain topography significantly affects the progression of wildland fires, with upslopes accelerating the rate of spread and downslopes decelerating it \citep{Byram1959}. One approach for incorporating the effect of inclination in physics-based fire spread models is by utilizing an effective wind speed, $\left<u_{eff}\right>$ ([=] m/s) \citep{Nelson2002}.The effective wind speed is determined by augmenting the air speed through the plantation with the local upstream component of the buoyant velocity of the thermal plume.
Assuming that the direction of wind coincides with the upslope--the latter characterized by an inclination $\theta$ with respect to the horizontal--the effective wind speed is defined as:
\begin{equation}
\label{eq:EffVel_1d} \left<u_{eff}\right>=\left<{u}\right> + \left<u_{buoy}\right>\sin{\theta}.
\end{equation}
The above expression can be readily generalized to the case when the wind direction forms an angle $\psi$ with the direction of upslope. Then, the local, total contribution is as follows:
\begin{equation}
\label{eq:EffVel_2d} \left<u_{eff}\right>=\left[\left(\left<{u}\right>\sin{\psi}\right)^2 + \left(\left<{u}\right>\cos{\psi} + \left<u_{buoy}\right>\sin{\theta}\right)^2 \right]^{1/2}.
\end{equation}

The following procedure is adopted for the computation of $\left<u_{eff}\right>$ \citep{Nelson2002}. The fireline intensity, $I_B$ ([=] W/m), which is a measure of the thermal power released per unit length of the firefront, is computed from the expression \citep{Byram1959}:
\begin{equation}
\label{eq:Intensity}
I_B = A_2 W_a ({\rm ROS}) = A_2 \alpha \rho_s H \left(S_{2,0} - S_{2,min}\right) ({\rm ROS})
\end{equation}
It is recalled that $A_2$ is the heat of combustion, $\alpha$ is the packing ratio, $\rho_s$ is the fuel density, and $H$ is the canopy height. Here, $W_a$ ([=] kg/m$^2$) is the mass of burned combustibles per unit area of the terrain (available fuel loading) and is estimated by subtracting from the initially available dry combustibles, $\alpha \rho_s H S_{2,0}$, the amount remaining after the passage of the firefront. A typical magnitude of buoyant velocity is then provided by the expression:
\begin{equation}
\label{eq:u_buoy0}
u_{buoy,0} = \left(\frac{2 g I_B}{\rho_g c_{pg} T_a}\right)^{1/3},
\end{equation}
with $g$ ([=] m/s$^2$) denoting the gravitational acceleration. The buoyant term is taken as representative of the vertical velocity at the location of maximum temperature, $T_{max}$, across the firefront. Thus, the buoyant velocity profile to be substituted in \cref{eq:EffVel_1d} is estimated as:
\begin{equation}
\label{eq:last_u_buoy0}
\left<u_{buoy}\right> = u_{buoy,0} \left(\frac{T - T_a}{T_{max} - T_a}\right)
\end{equation}

\subsection{Final form and numerical implementation of the model}
\label{sec:Equations_Numerical}

The following are the final equations of the model: 
\begin{align}
\label{eq:EnergyBalanceDimlessModel}
\frac{\partial T}{\partial t} &= \frac{c_1}{c_0}\left(\underbrace{\mathcal D_{\rm eff,x}\frac{\partial^2 T}{\partial x^2} + \mathcal D_{\rm eff,y}\frac{\partial^2 T}{\partial y^2}}_{\text{dispersion}}-\underbrace{\left<u_{eff,x}\right> \frac{\partial T}{\partial x} - \left<u_{eff,y}\right>\frac{\partial T}{\partial y}}_{\text{advection}} \right)-\underbrace{
\frac{c_2}{c_0}\,S_1\,r_1+\frac{c_3}{c_0}\,S_2\,r_{2t}}_{\text{reaction}}-\underbrace{
\frac{c_4}{c_0}\,U(T-T_a)}_{\text{convection}}, \\
\label{eq:Reaction1Model}
\frac{\partial S_1}{\partial t} &=-S_1\,r_1, \\
\label{eq:Reaction2Model}
\frac{\partial S_2}{\partial t} &=-S_2\,r_{2t}.
\end{align}
The coefficients $c_i$, which are combinations of thermo-physical properties, are given below: 
\begin{align}
c_0&=\alpha S+(1-\alpha)\lambda\gamma+\alpha\gamma (1-S), \\
c_1&=(1-\alpha)\lambda\gamma+\alpha\gamma (1-S)=c_0-\alpha S,\\
c_2&= \alpha \, \frac{A_1}{c_{ps}},\\
c_3&= \alpha \, \frac{A_2}{c_{ps}},\\
c_4&= \frac{1}{H \rho_s c_{ps}}, 
\end{align} 
where $\gamma= c_{pg}/c_{ps}$, and it is recalled that $\lambda= \rho_g/\rho_s$.

The set of eqs.~\eqref{eq:EnergyBalanceDimlessModel}, \eqref{eq:Reaction1Model}, and \eqref{eq:Reaction2Model} is discretized by an explicit, finite difference numerical scheme on a uniform, rectangular grid. First-order upwinding is used for the advection terms and second-order central differences for the dispersion terms. The resulting system of temporal ODEs is solved by Adams–Bashforth methods \citep{Hairer} with a time step that ensures both convection and dispersion stability \citep{RaimundBürger2020}.
A localized one- or two-dimensional temperature spike of Gaussian shape is used as the initial condition, and open outflow boundary conditions are implemented at the downstream boundaries in order to allow the firefront to move smoothly out of the computational domain without significant backward influence \citep{Papanastasiou1992, Karniadakis2014}. (These conditions essentially amount to extending the validity of the discretized equations at the boundary nodes using appropriate one-sided approximations for the spatial derivatives, and their effectiveness is confirmed by the very good agreement with representative runs on a computational domain twice the original size.)

The two-dimensional simulations to be discussed in \Cref{sec:FirePatterns} and \Cref{sec:2D_Case_Studies} refer to a 500 m x 500 m domain and use grid spacing $\Delta x=\Delta y= 0.5$ m and typical time step $\Delta t=0.1$ s. These runs were performed in Matlab\textregistered\; environment on a single-processor AMD Ryzen Threadripper 3995WX computer and took 3 s of computation time per 1 s of simulated time. However, an adaptive mesh refinement (AMR) technique \cite{Chatzimanolakis2022}, which is particularly appropriate for the present problem, is presently in the process of implementation. The methodology is outlined below and it is reported that preliminary re-runs on an 8-core computer took only 0.1 s of computation time per 1 s of simulated time.

The AMR technique is implemented by the CubismAMR library \cite{ Chatzimanolakis2022a}, which discretizes the computational domain into square blocks of locally uniform resolution.  At the interfaces between adjacent blocks of differing resolutions, CubismAMR performs the necessary interpolations creating a uniform resolution frame around each grid point. This allows the use of finite difference schemes designed for uniform grids. Mesh refinement is driven by the temperature gradient; if either \(\frac{\partial T}{\partial x}\) or \(\frac{\partial T}{\partial y}\) exceeds a user-defined upper threshold (set to \(10 \, \text{K/m}\)), the mesh is refined. Conversely, the mesh is compressed when the gradient falls below a lower threshold (set to \(0.1 \, \text{K/m}\)).

\section{Model validation and results}
\label{sec:Validation}

\subsection{A representative one-dimensional simulation}
\label{sec:1D_Case}

Having developed a model for wildfire spread, the next objective is to probe its behavior and compare its predictions to observations that are known from laboratory experiments and field studies. First, the model's one-dimensional (1D) version is used to examine the effects of various fuel properties, including bulk density, moisture content, particle size, and also the combined effect of wind speed and terrain inclination. The 1D version corresponds to a very long fireline that moves uniformly in the streamwise ($x-$) direction. As it is known that the rate of spread, $\rm ROS$, increases with the length $w$ of the firefront \citep{Cheney1993,Linn2005}, the 1D model is expected to provide an upper limit to the actual $\rm ROS$.

\begin{figure}[tbp]
\centering
\begin{tabular}{cc}
\includegraphics[width=0.45\textwidth]{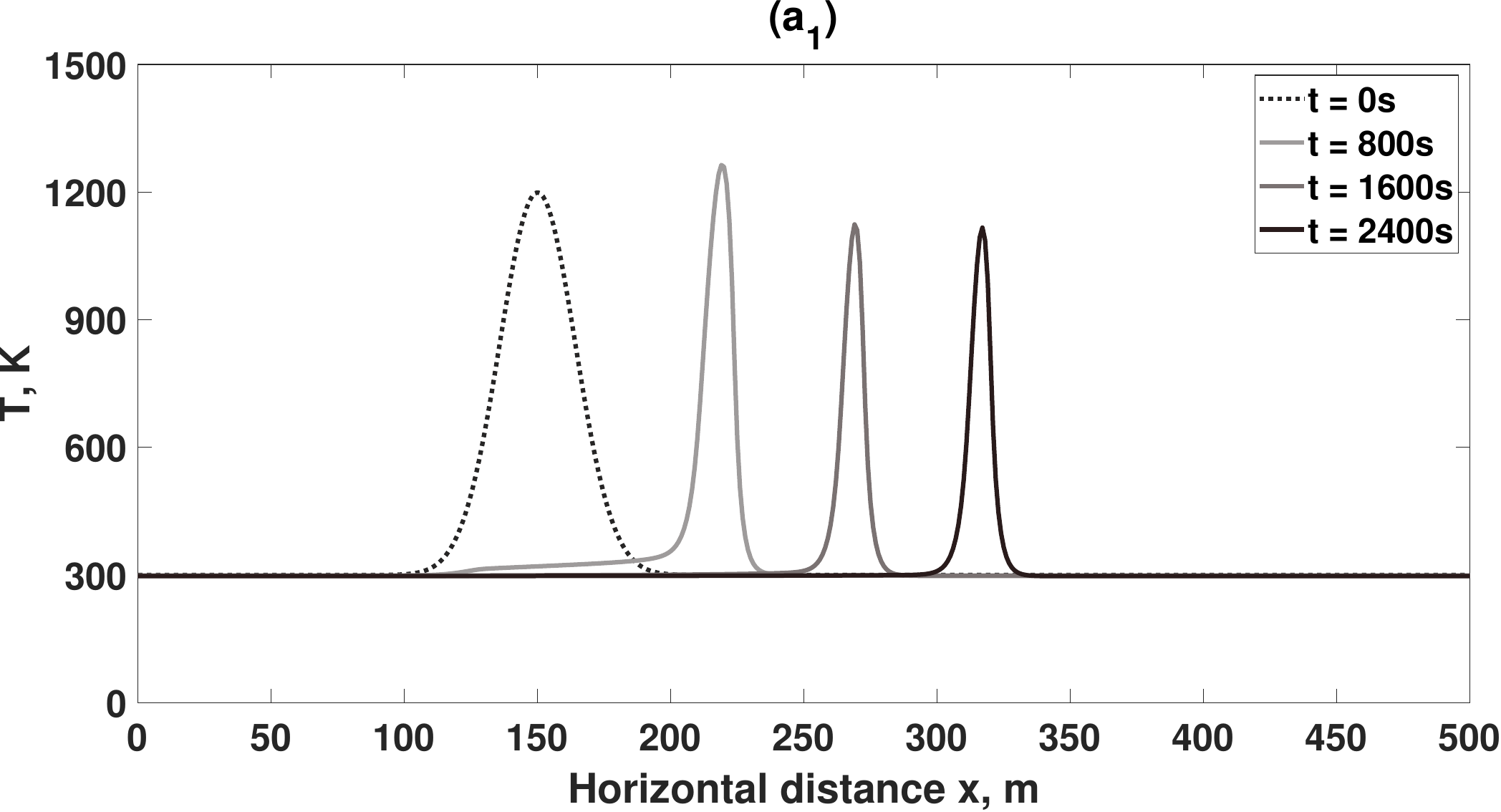
} &
\includegraphics[width=0.45
\textwidth]{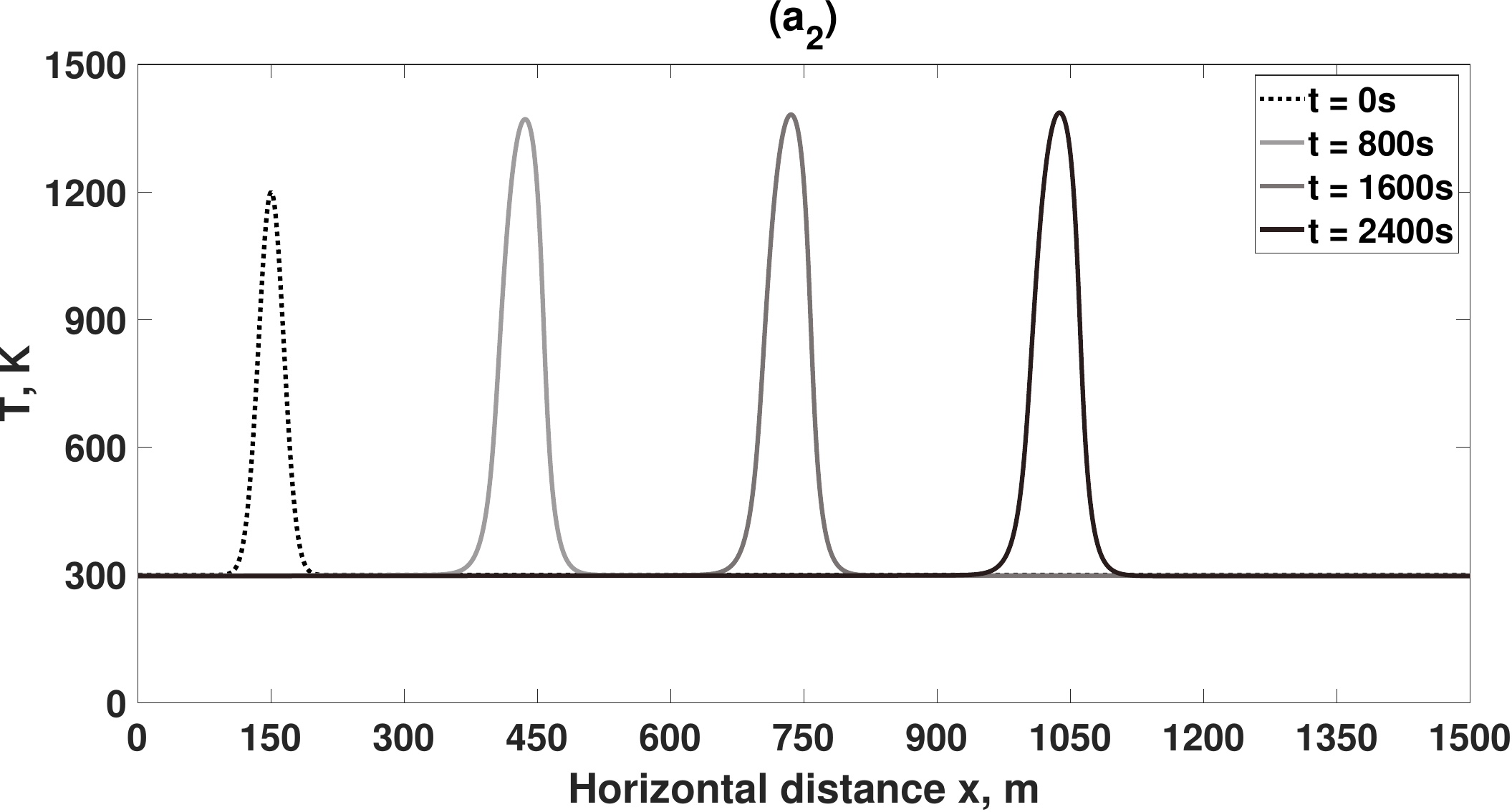
}\\ 
\includegraphics[width=0.45
\textwidth]{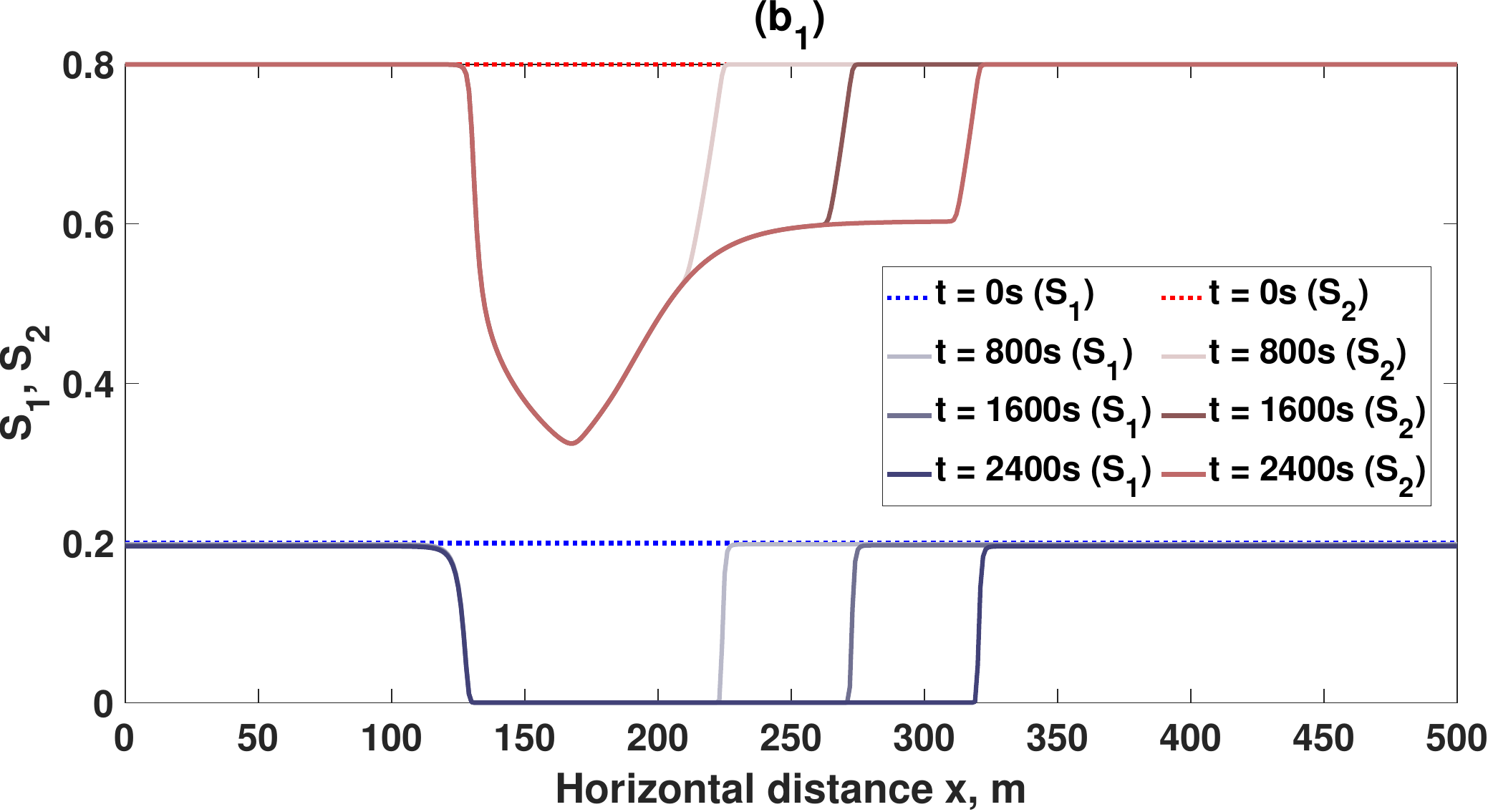
} &
\includegraphics[width=0.45\textwidth]{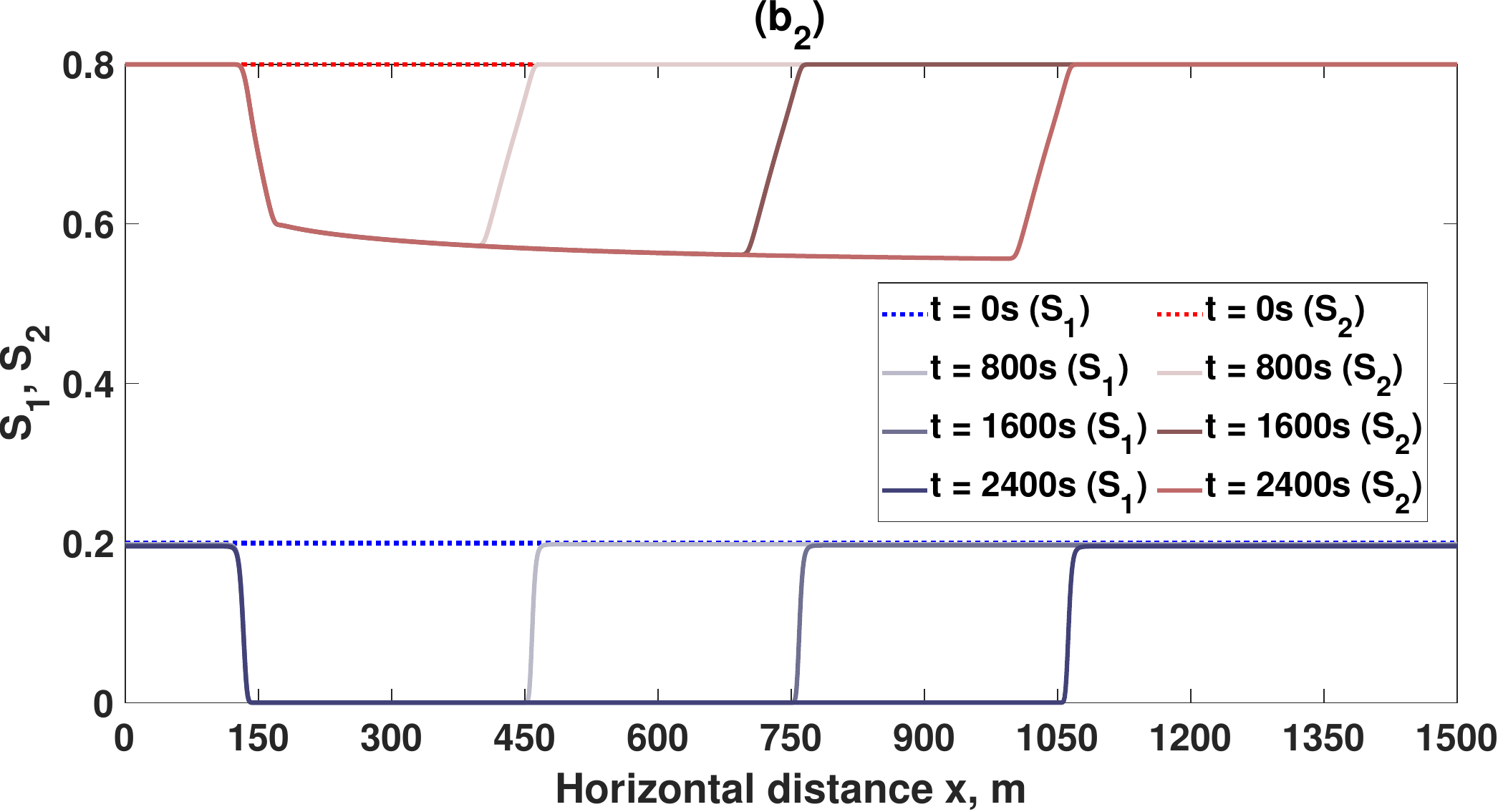
} \\
\includegraphics[width=0.45\textwidth]{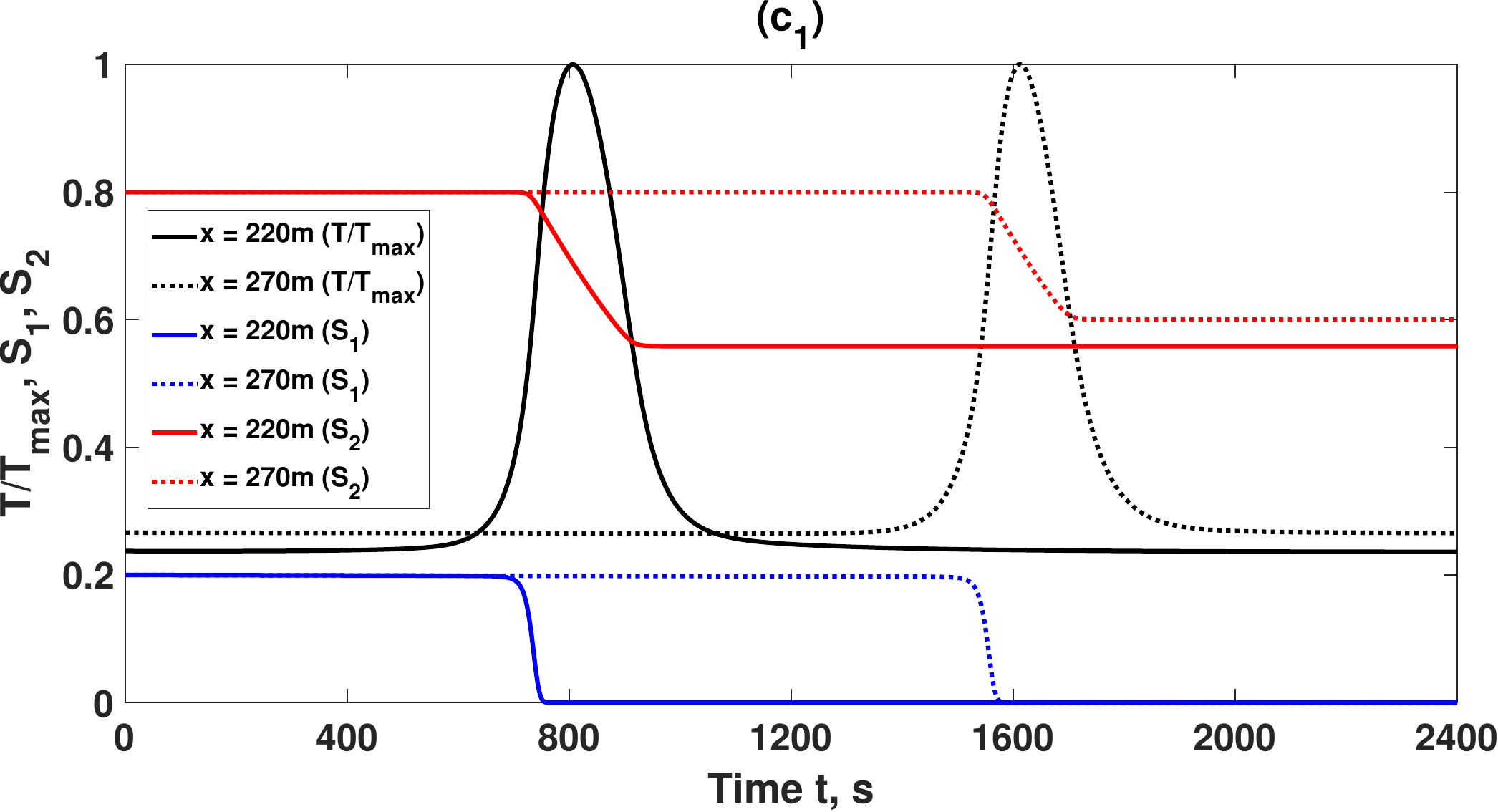} &
\includegraphics[width=0.45\textwidth]{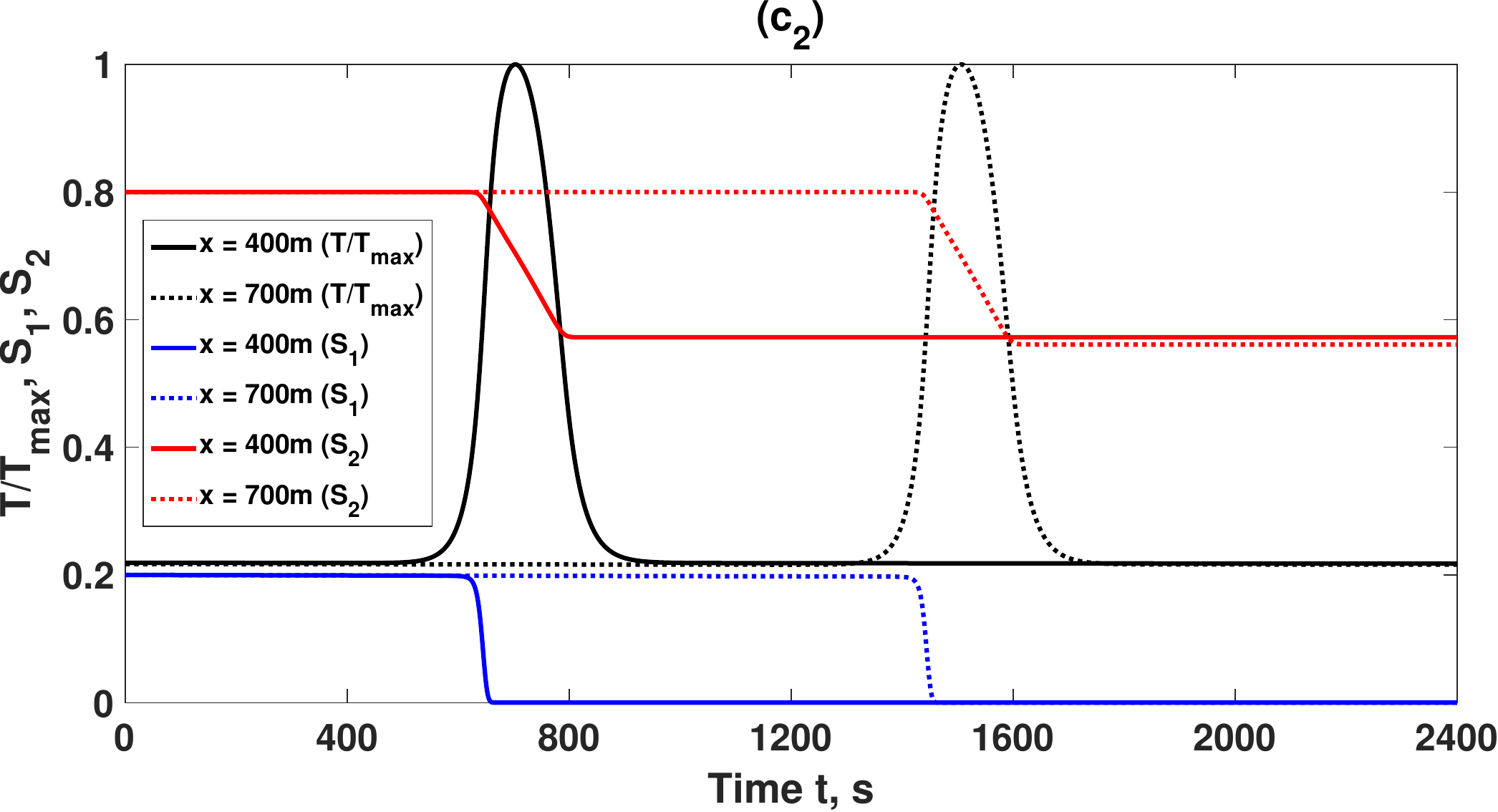} 
\end{tabular}
\caption{(a) The spatial distribution of temperature at time instants $t=$ 0, 800, 1600 and 2400 s, (b) the spatial distribution of $S_1$ and $S_2$ at the same time instants, (c) the temporal variation of dimensionless temperature $T/T_{\max}$, $S_1$ and $S_2$. The subscript 1  (left column) is for $u_{10}=$ 3 m/s and the subscript 2 (right column) for $u_{10}=$ 10 m/s, while the maximum localized ignition point is everywhere at $x_0=$ 150 m from the origin.}
\label{fig:ReprRes}
\end{figure}

The results for a representative case are shown in \cref{fig:ReprRes}, which corresponds to the parameter values listed in \cref{table:ModelParameters} and the values for a dense canopy in \cref{table:WinsSpeedParameters}. \Cref{fig:ReprRes}(a) depicts the spatial variation of temperature for (a$_1$) $u_{10}=$ 3 m/s and (a$_2$) $u_{10}=$ 10 m/s at time instants, $t=$ 0, 800, 1600 and 2400 s, from the onset of the initial spikes (dashed lines). The progression of the firefront with time is evident (solid lines), and the $\rm ROS$ is readily calculated from the displacement of the temperature maximum and the corresponding time lag. Such a calculation indicates that sometimes the $\rm ROS$ increases gradually with time, a behavior reminiscent of fire's acceleration with size. Such a concept has been included in some prediction models, such as FARSITE or FIRETEC \citep{Pimont2012}. Aiming to overcome possible ambiguity, the $\rm ROS$ reported in the parametric investigations of this section is the mean value calculated from the crest progression between 1000 s and 1500 s from ignition.

\begin{table}[tbp]
\centering
\caption{Representative values of the model parameters.}
\label{table:ModelParameters}
\begin{tabular}{ll|ll|ll|ll}
\toprule
\textbf{Parameter} & \textbf{Value} & \textbf{Parameter} & \textbf{Value} & \textbf{Parameter} & \textbf{Value} & \textbf{Parameter} & \textbf{Value} \\
\midrule
$\rm FMC$ & 25\% & $c_{pg}$ & 1043 & $A_1$ & 22$\cdot10^5$ & $\sigma$ & 20 \\
$H$ & 2 & $c_{ps}$ & 1800 & $A_2$ & 2$\cdot10^7$ & $A_{nc}$ & 0.2 \\
$T_a$ & 300 & $c_{s1}$ & 30 & $D_{\rm rb}$ & 0.1 & $A_d$ & 0.125 \\
$T_{max,i}$ & 1200 & $c_{s2}$ & 40 & $r_{m,0}$ & 0.002 & $\eta$ & 3 \\
$\rho_g$ & 1 & $b_1$ & 4500 & $r_{m,c}$ & 0.004 & $\alpha$ & 0.002 \\
$\rho_s$ & 700 & $b_2$ & 7000 & $\gamma_d$ & 0.03 & $\epsilon$ & 0.2 \\
\bottomrule
\end{tabular}
\end{table}

\Cref{fig:ReprRes}(b) depicts the spatial distribution of $S_1$ and $S_2$, i.e., the dimensionless mass of water and combustibles remaining in the solid phase, for (b$_1$) $u_{10}=$ 3 m/s and (b$_2$) $u_{10}=$ 10 m/s at the same time instants, $t=$ 0, 800, 1600, and 2400 s from ignition. As expected, water has totally evaporated everywhere the flame has passed (blue-type lines) before the exothermic reaction takes place (red-type lines). However, varying amounts of combustible solids remain, depending on how fast the combustion proceeds and how efficiently the burned area is cooled down by the incoming wind. The increase in consumed combustibles with downstream distance explains the acceleration of the $\rm ROS$ that is sometimes observed: as fire moves ahead, it intensifies and burns more fuel during its passage, thus the firefront spreads faster. However, a steady-progressive state evidently establishes further downstream, which corresponds to an asymptotic limit $S_2\rightarrow S_{2,min}$.

As an alternative representation of the same process, \cref{fig:ReprRes}(c) illustrates the three variables, $T/T_{max}$, $S_1$ and $S_2$, as functions evolving over time, for (c$_1$) $u_{10}=$ 3 m/s and (c$_2$) $u_{10}=$ 10 m/s with $t \in$ [0, 2400] s, at two fixed spatial locations, x = 220, 270 m and x = 400, 700 m, respectively, from the ignition spot (term $T_{\max}$ is the maximum temperature achieved at each location over time). The onset of combustion with the arrival of the firefront is evident, as is the temperature decline behind the front, depicted by a tail that may be longer or shorter depending mainly on the wind speed and terrain inclination. The remaining amount of combustibles, $S_2$, stabilizes to a constant value after some time, indicating extinction below the minimum ignition temperature. Again, it is evident that the remaining combustibles may gradually decrease with downstream distance in case the fire intensifies.

\subsection{Fuel modeling flexibility: bulk density, moisture content and particle size}
\label{sec:FuelCharacteristics}

The model offers a number of parameters that may be combined to provide a realistic representation of the fuel's properties. The bulk density of solid material, defined as $m_{s,0}=\alpha\rho_s$, in terms of the volume fraction $\alpha$, and the material density $\rho_s$, is known to have a systematic effect on the fire spread rate. More specifically, it has been observed \citep{Thomas1971, Carrier1991, Wolff1991, Catchpole1998} that increasing bulk density leads to slower $\rm ROS$. This behavior has been mathematically expressed by an inverse power law, $\rm ROS$ $\sim m_{s,0}^{-\zeta}$, which appears to satisfactorily describe both laboratory experiments and field observations \citep{Vega1998, Fernandes2001, Marino2012, Thomas1971, Carrier1991, Wolff1991, Catchpole1998}. However, the proposed exponent varies widely among the above studies, moving in the range [0.23, 0.74].

\begin{figure}[tbp]
\centering
\includegraphics[width=.55\textwidth]{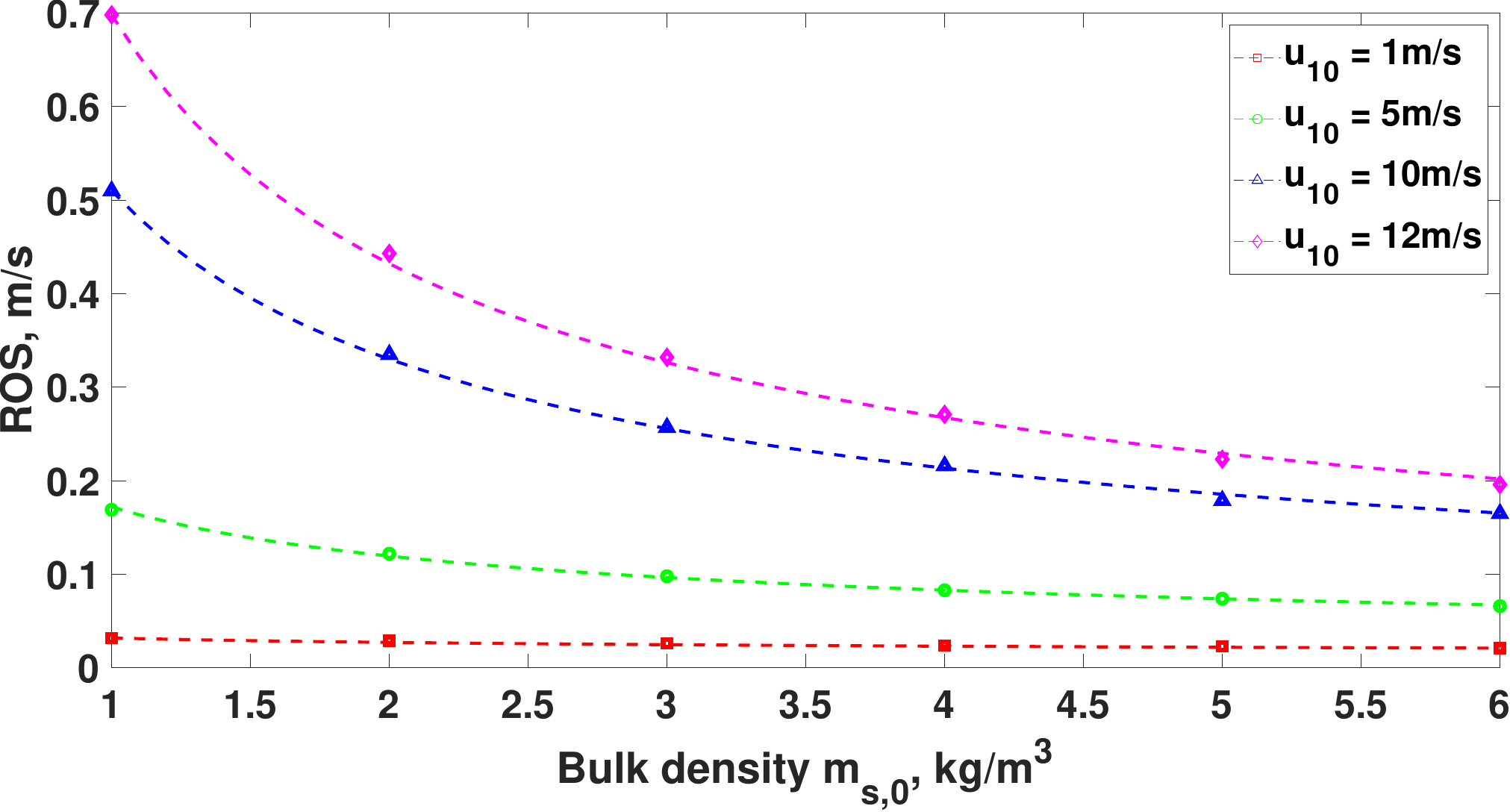}
\caption{The $\rm ROS$ as a function of the fuel bulk density, $m_{s,0}$, for wind speeds, $u_{10}=$ 1, 5, 10, and 12 m/s. Points are simulation results and lines the best-fit inverse power law.}   \label{fig:BulkDensity}
\end{figure}

The present model confirms the central role of bulk density. In particular, keeping all other parameters constant as outlined in tables~\ref{table:WinsSpeedParameters}, \ref{table:ModelParameters}, the $\rm ROS$ is found to vary only with $m_{s,0}$ and not with $\alpha$ or $\rho_s$ independently. The inverse power law is also followed by the model, with the exponent, $\zeta$, actually being an increasing function of the air velocity, $u_{10}$, above the canopy. Indicative examples for $u_{10}=$ 1, 5, 10, and 12 m/s, and $m_{s,0}$ in the range [1, 6] kg/m$^3$, are shown in \cref{fig:BulkDensity}.
The exponent $\zeta$ appears in our model to follow the empirical fit, $\zeta=0.04\, (1+u_{10})$, with very good accuracy.

The increase in fuel moisture content is also known to result in slower $\rm ROS$ \citep{Marino2012, Rossa2017} and to eventually lead to the extinction of the flame under sufficiently wet conditions. An exponential functionality of the form $\rm ROS$ $\sim e^{-\mu\, (FMC)}$ has been repeatedly proposed \citep{Cheney1993, Fernandes2009, Marino2012, KhalidMoinuddin2021}, but the exponent $\mu$ is found to vary widely. In particular, it is higher when the correlation is based on dead moisture content and lower when using the mean value of dead and live moisture. Predictions of the present model, which are based on the total moisture content, are satisfactorily correlated by the same functional form and values of the exponent $\mu$ in the range $[0.013,0.017]$, or even smaller for very low air velocities. The FMC that results in fire extinction is predicted to increase with bulk density and decrease with air velocity, i.e., a fire on wet fuel is more persistent at high bulk density and low air speed.

The effect of fuel particle size is also critical, as leaves and small-diameter sticks have larger surface-to-volume ratios and thus burn faster than thick branches. This behavior may be quantified by the appropriate choice of the constants in the combustion kinetics. With reference to \cref{eq:Reaction2}, a characteristic burning time may be defined as:
\begin{equation}
\label{eq:BurningTime}
 t_c = -\frac{\ln(S_{2,f}/S_{2,0})}{c_{s2}\,e^{-b_2/T_c}},
\end{equation}
in terms of the initial exothermic composition, $S_{2,0}$, and the remaining, $S_{2,f}$, combustible material after time $t_c$. Taking this ratio as equal to 0.1 and using a characteristic burning temperature, $T_c \approx$ 1250 K \citep{Taylor2004}, gives the estimate $t_c\approx 1000/c_{s2}$. Values of $c_{s2}$ in the range $[5, 200]$ s$^{-1}$ result in $t_c$ in the range $[200,5]$ s. This range is very favorably compared with the fuel residence times for a variety of fuels in the extensive database presented in \citep{Nelson1988}.

\subsection{The role of wind and terrain inclination}
\label{sec:Wind_Slope}

As evidenced already, the overlying wind speed is a key ingredient in determining the fire propagation speed and direction \citep{Weise1997, Pimont2012, Simeoni2001, Taylor2004, Banerjee2020}. In the absence of wind, the fire is expected to spread symmetrically, with heat for ignition being transported by buoyant dynamics and radiation, as modeled by the term $D_{\rm rb}$, of the effective diffusivity. With increasing wind speed, fire propagation is accelerated in the direction of the wind and decelerated against it. Beyond a value of wind speed, propagation against the wind is arrested, and finally the backward front extinguishes \citep{Gao2021}.

\begin{figure}[tbp]
\centering
\includegraphics[width=0.45\textwidth]{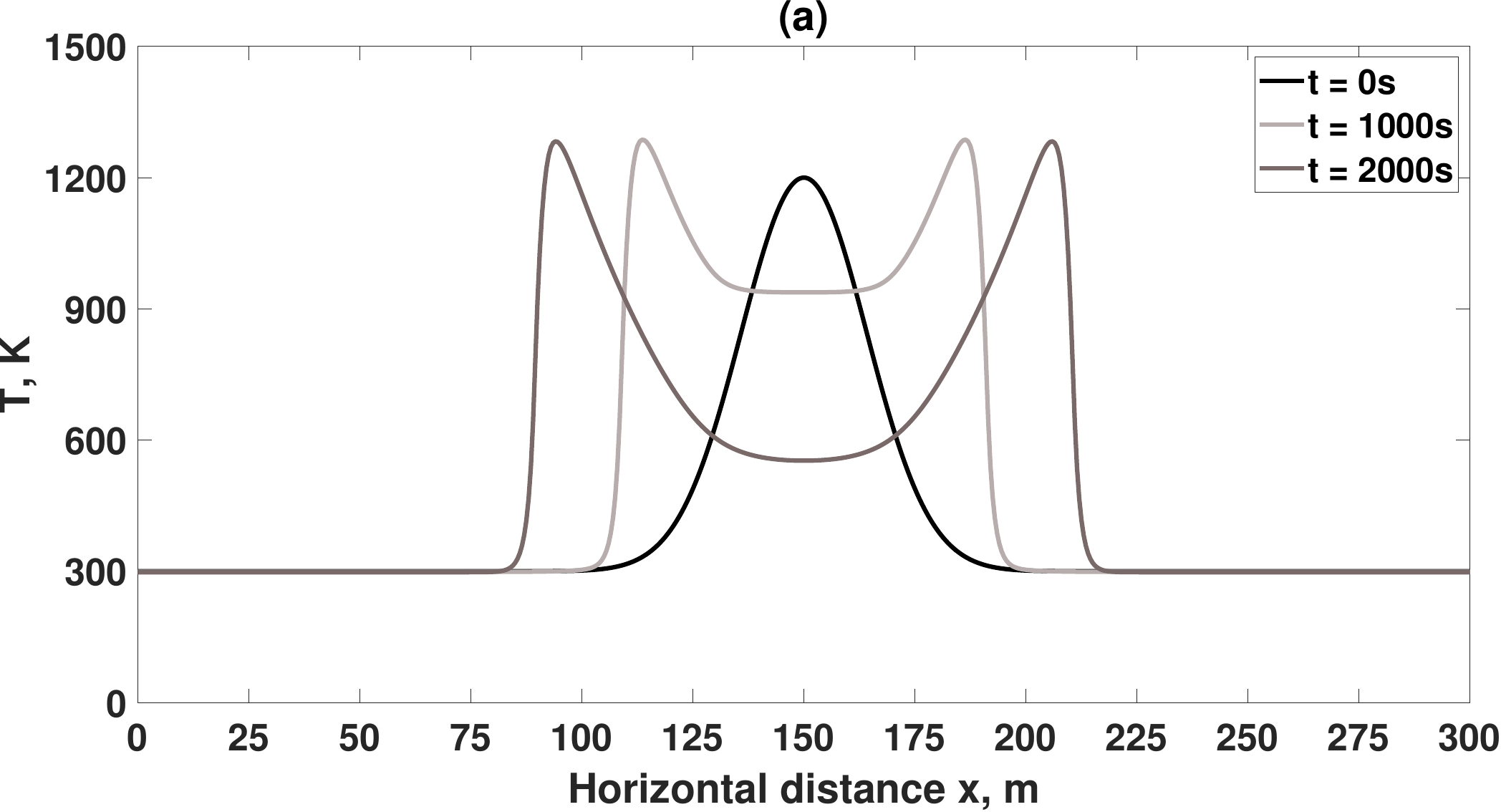} \includegraphics[width=0.45\textwidth]{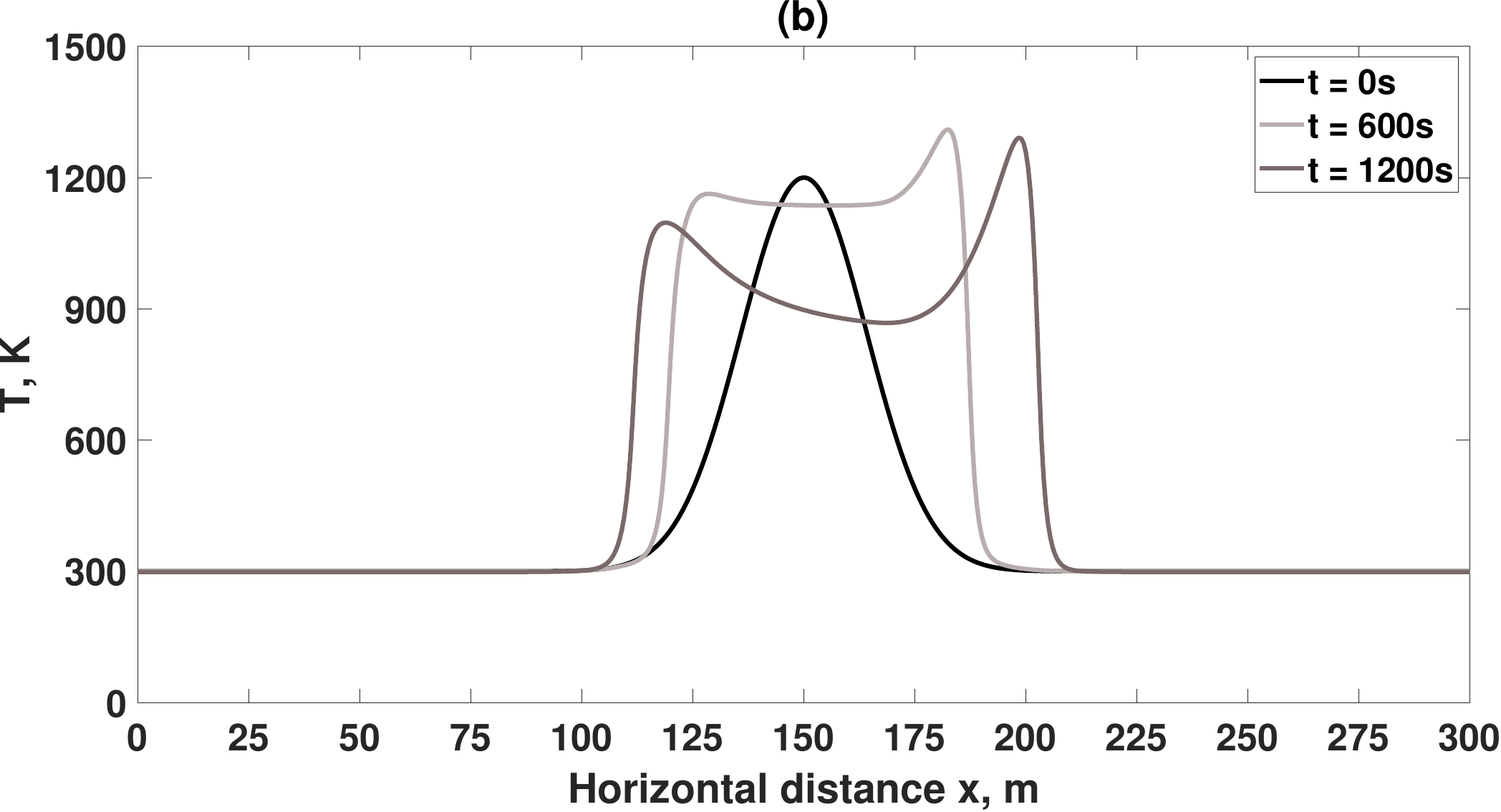} \includegraphics[width=0.45\textwidth]{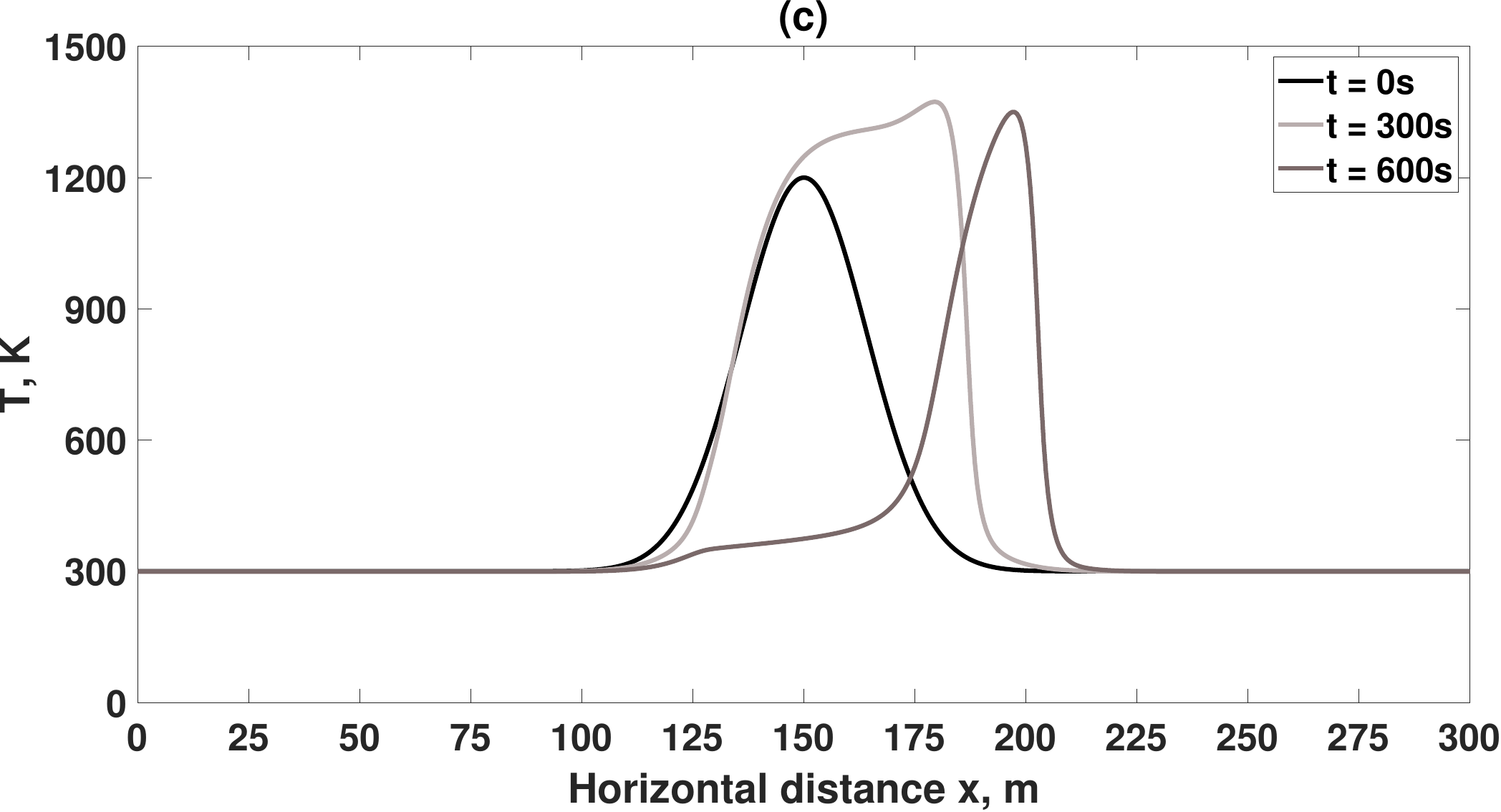}
\caption{The spatial distribution of temperature at different time instants separated by $\Delta t_{s}$ seconds. (a) $u_{10}=$ 0 m/s, $\Delta t_{s}=$ 1000 s, (b) $u_{10}=$ 1 m/s, $\Delta t_{s}=$ 600 s and (c) $u_{10}=$ 2.5 m/s, $\Delta t_{s}=$ 300 s.}
\label{fig:LowSpeed}
\end{figure}

In the proposed model, the effect of wind is introduced by a local mean velocity of the gaseous phase inside the plantation, as described in \Cref{sec:WindSpeed}. Representative examples of front propagation are shown in \cref{fig:LowSpeed}. Specifically, \cref{fig:LowSpeed}(a) corresponds to zero wind speed, $u_{10}=$ 0 m/s, and exhibits symmetric propagation in both directions, as evidenced by the temperature profiles for time instants, $t=$ 0, 1000, and 2000 s. Introducing a small wind speed, $u_{10}=$ 1 m/s, in \cref{fig:LowSpeed}(b) results in the acceleration of the wavefront moving with the wind and the deceleration of the front moving against the wind, as evidenced by the temperature profiles, now for time instants, at $t=$ 0, 600, and 1200 s. Increasing the wind speed further enhances the difference between the spread rates of the two fronts. Eventually, beyond $u_{10}=$ 2.3 m/s, the wind-opposed front blows off and the fire spreads only in the direction of the wind (see \cref{fig:LowSpeed}(c) for $u_{10}=$ 2.5 m/s and $t=$ 0, 300, and 600 s).

\begin{figure}[tbp]
\centering  \includegraphics[width=.5\textwidth]{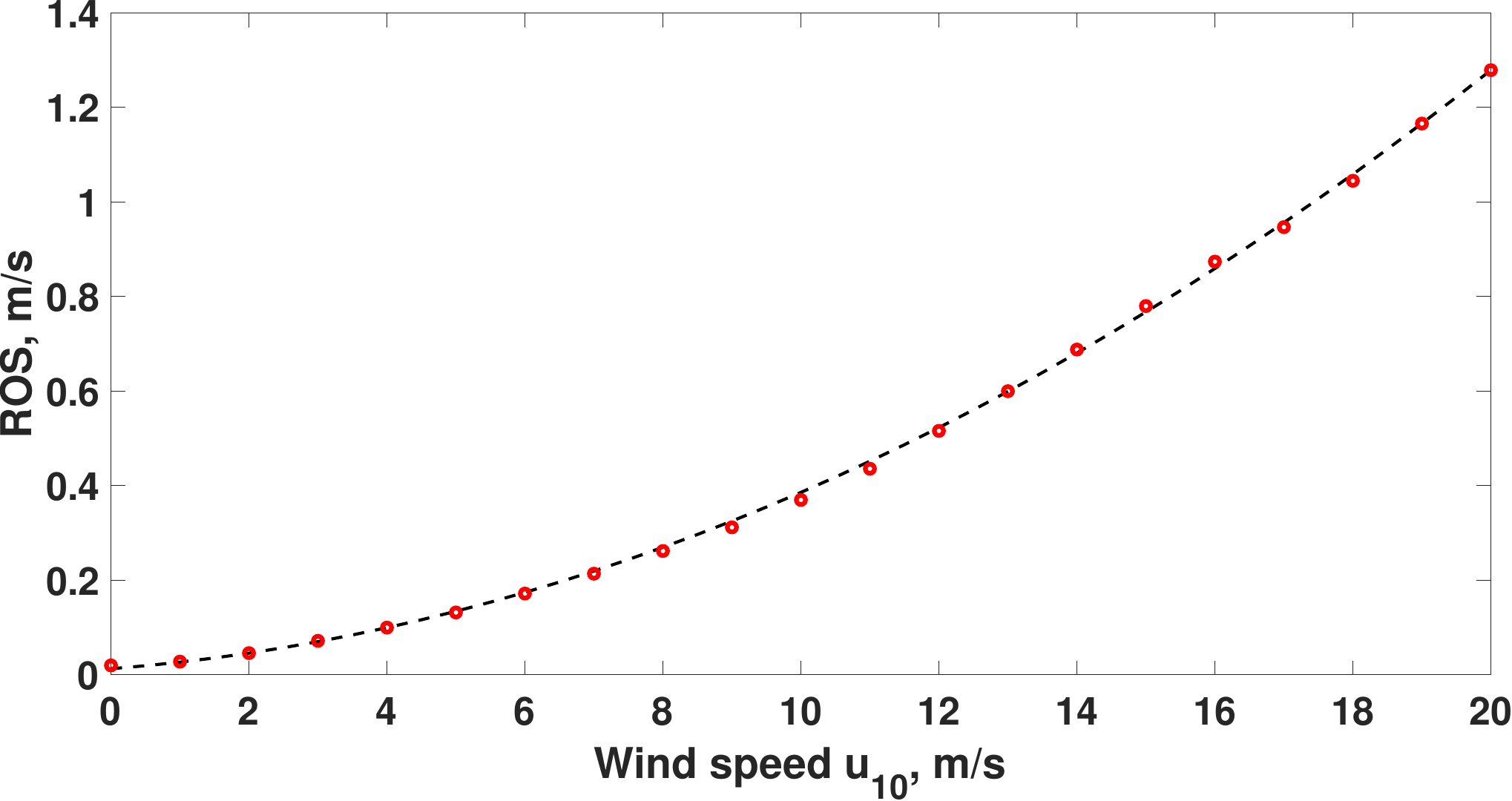}
\caption{The ${\rm ROS}$ as function of the wind speed $u_{10}$ with all other parameters as listed in \cref{table:ModelParameters}. Points are simulation results and line the best-fit parabolic curve.}
\label{fig:WindSpeed}
\end{figure}

The predicted dependence of the steady $\rm ROS$ on wind speed $u_{10}$ is depicted quantitatively in \cref{fig:WindSpeed}. Apart from the variation of $u_{10}$, all other parameter values are listed in \cref{table:ModelParameters}. It is noted that the numerical results in \cref{fig:WindSpeed} follow very closely a parabolic curve, ${\rm ROS}={\rm (ROS)}_0+\Phi u_{10}^2$, with ${\rm (ROS)}_0$ the rate of spread at zero wind speed, while the coefficient $\Phi$ is evidently expected to vary with the fuel properties.

\begin{figure}[tbp]
\centering  \includegraphics[width=.5\textwidth]{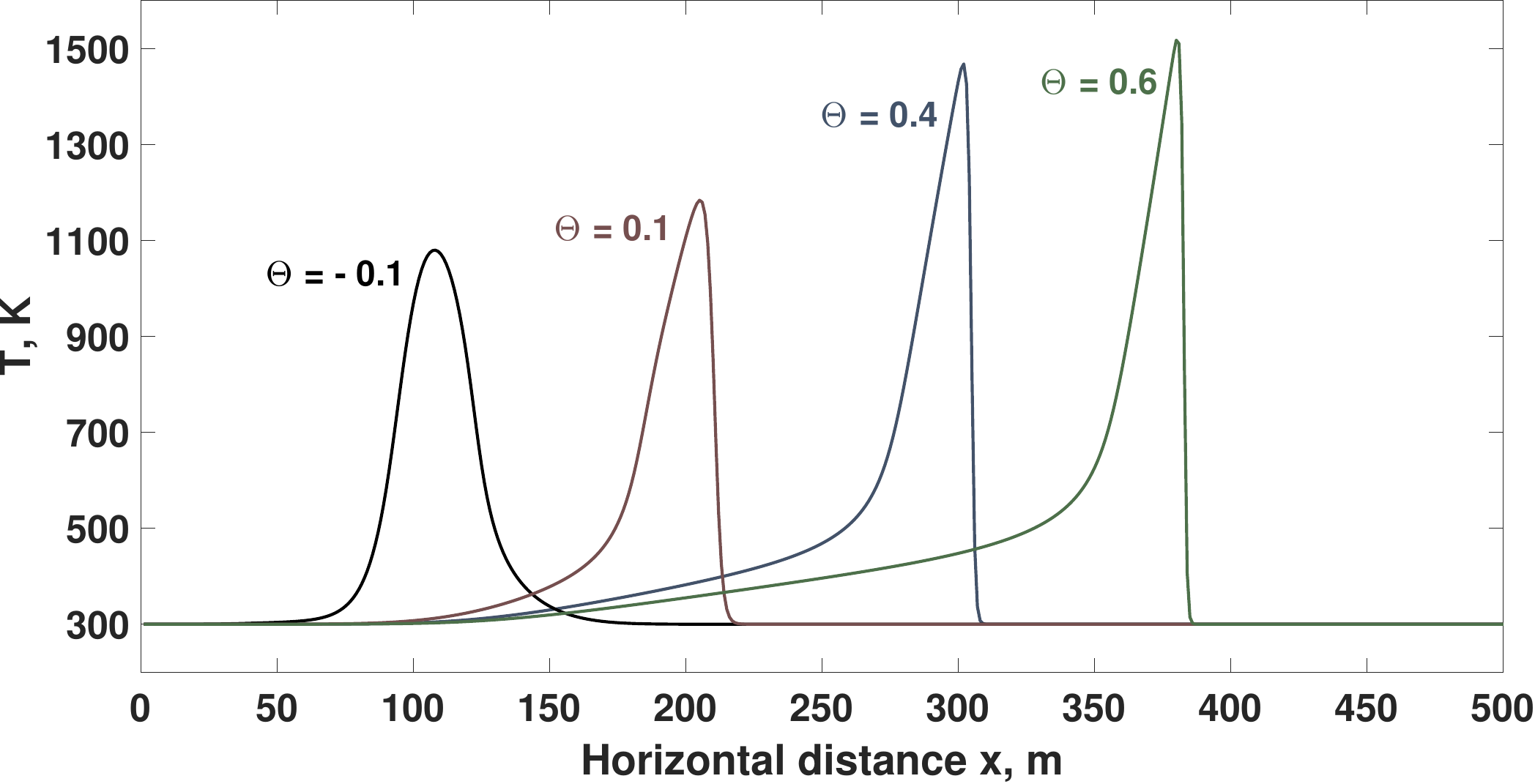}
\caption{The temperature profile of the firefront, 500 s after ignition at $x=$ 50 m, for $\Theta=$ - 0.1, 0.1, 0.4, and 0.6.}
\label{fig:Inclined_1}
\end{figure}

When the terrain is flat but at an inclination $\theta$ with respect to the horizontal, the wind velocity through the canopy is substituted by the enhanced distribution $\left<u_{eff}\right>$, \cref{eq:EffVel_1d}, which is computed by the methodology described in \cref{sec:Inclination}. Following a literature convention \citep{Pimont2012}, the inclination is presently characterized by the value $\Theta=\tan{\theta}$. The change in the location and temperature profile of the firefront for terrain inclinations corresponding to $\Theta=$ - 0.1, 0.1, 0.4, and 0.6 is shown in \cref{fig:Inclined_1}. All curves represent the condition after a lapse a of 500 s from ignition at $x=$ 50 m. The first curve ($\Theta=$ - 0.1) corresponds to a downslope, and the front is roughly symmetric and spreading slowly. With increasing upslope (positive values of $\Theta$), the front moves faster and becomes distinctly steeper, while the high-temperature tail is gradually elongated.

However, the key variable that needs to be predicted is--as always--the rate of spread. In order to test the performance of the present model for inclined terrain, its predictions are compared to the results from three-dimensional (3D) CFD simulations \citep{Pimont2012} and from the frequently used semi-empirical operational model by Rothermel \citep{Rothermel1972, Andrews2018}. The latter is based on the rule:
\begin{equation}
\label{eq:Ros_Roth}
{\rm (ROS)} = {\rm (ROS)}_0 \left( 1 + \phi_U + \phi_{\Theta} \right),
\end{equation}
where the contribution of inclination angle $\theta$ is given by:
\begin{equation}
\label{eq:Inclination_Roth}
\phi_{\Theta} = \pm \frac{5,275}{\alpha^{0.3}}\,\tan^2{\theta} = \pm \frac{5,275\,\Theta ^2}{\alpha^{0.3}},
\end{equation}
with the minus sign used for downslope. Following Pimont et al. \citep{Pimont2012}, the rate of spread in still air was taken as ${\rm (ROS)}_0=$ 0.03 m/s, and $\phi_U$ was chosen so as to produce, for zero inclination, the same ${\rm (ROS)}$ as in the CFD simulations. For the same reason, the packing ratio in our simulations was varied slightly between different wind speeds.

\begin{figure}[tbp]
\centering  \includegraphics[width=.45\textwidth]{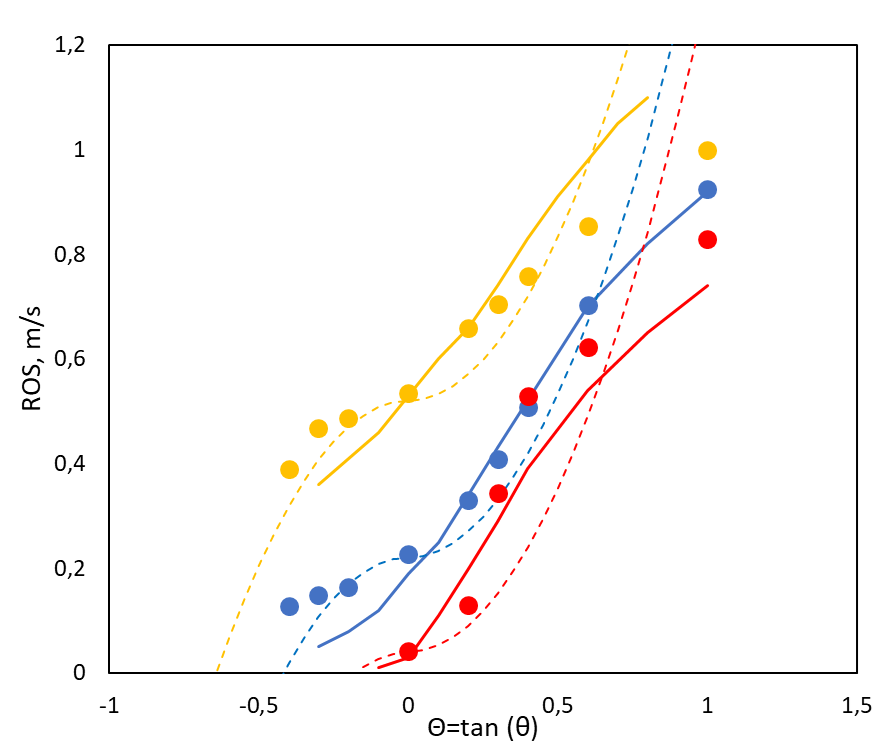}
\caption{The rate of spread of the firefront for various terrain inclinations and wind speed 1 m/s (red), 5 m/s (blue) and 12 m/s (yellow). Continuous lines are predictions of the present model, points are from detailed 3D simulations \citep{Pimont2012} and dashed lines are from the semi-empirical model of Rothermel \cite{Rothermel1972, Andrews2018}.}
\label{fig:Inclined_2}
\end{figure}

The results of the comparison, shown in \cref{fig:Inclined_2}, are very favorable for the present model. The combined effect of wind and inclination leads to variations in ${\rm ROS}$ that follow closely those of the detailed 3D simulations. There is a small over-prediction at the highest wind speed  and a small under-prediction at the lowest, with the results for $u_{10}=$ 5 m/s in surprising agreement. In particular, it is notable that the present predictions (similar to the CFD numeric) show a rapid increase in the $\rm{ROS}$ with inclination, while the semi-empirical rule predicts a slower increase at small inclinations and unrealistically high values at high inclinations. Thus, the present model appears to offer predictions similar to those of detailed 3D simulations at a small fraction of the computational time (0.1 s of computation time on a single-processor laptop per 1 s of simulation time).

\subsection{Two-dimensional wildfire dynamics: Evolution from a localized ignition site}
\label{sec:FirePatterns}

\begin{figure}[tbp]
\centering
\begin{tabular}{ccc}
\includegraphics[width=0.343\textwidth]{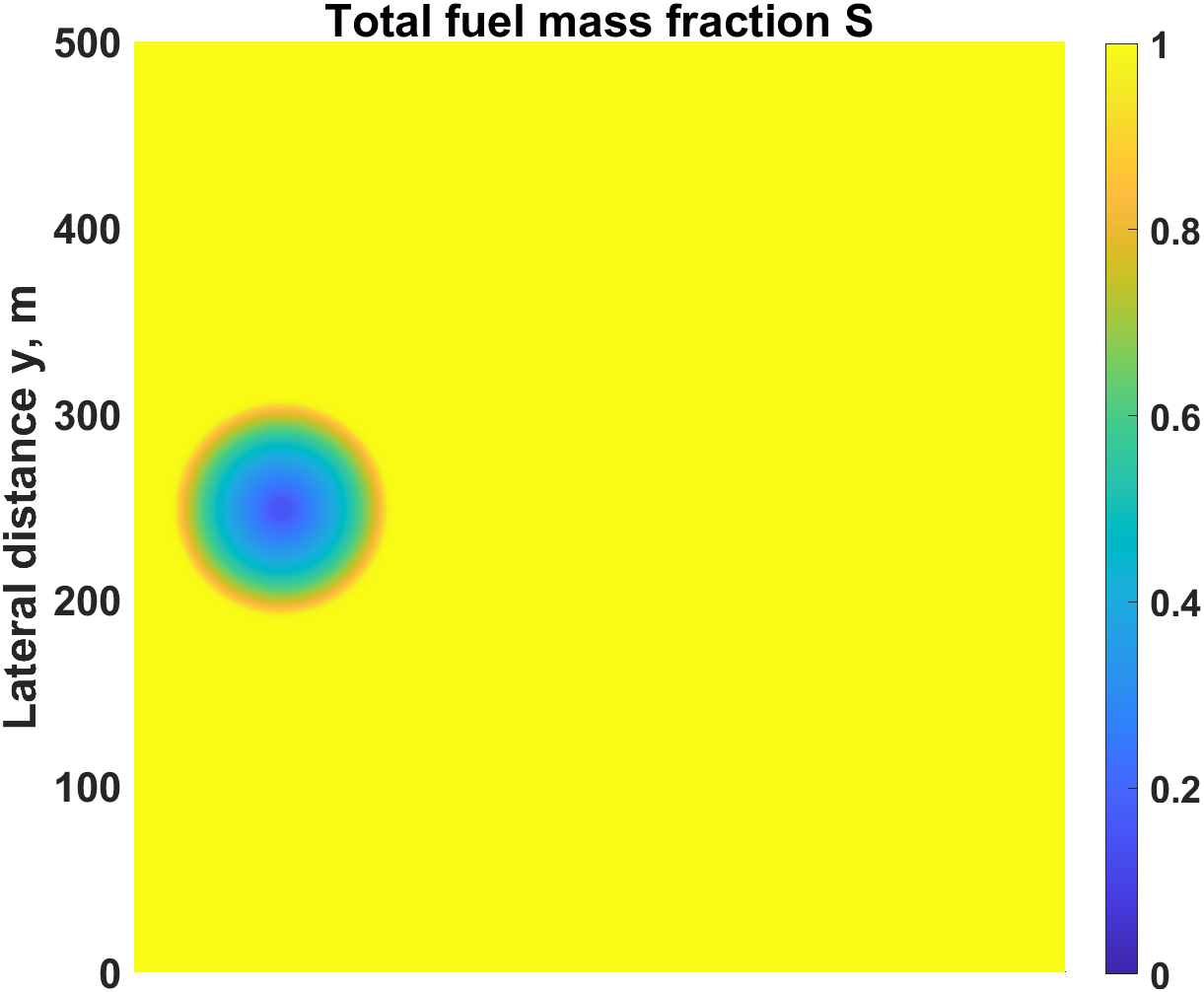} & &
\includegraphics[width=0.315\textwidth]{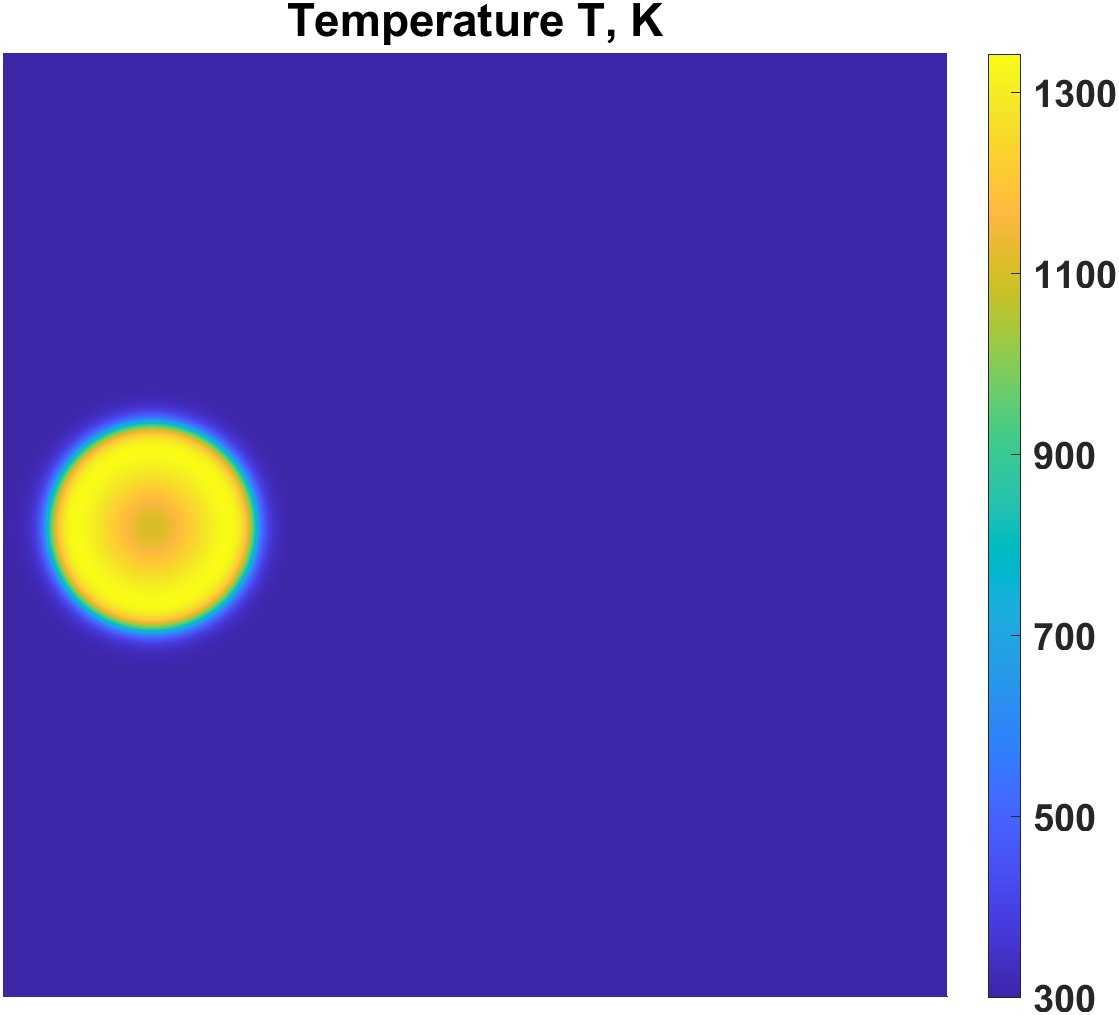} \\
\includegraphics[width=0.343\textwidth]{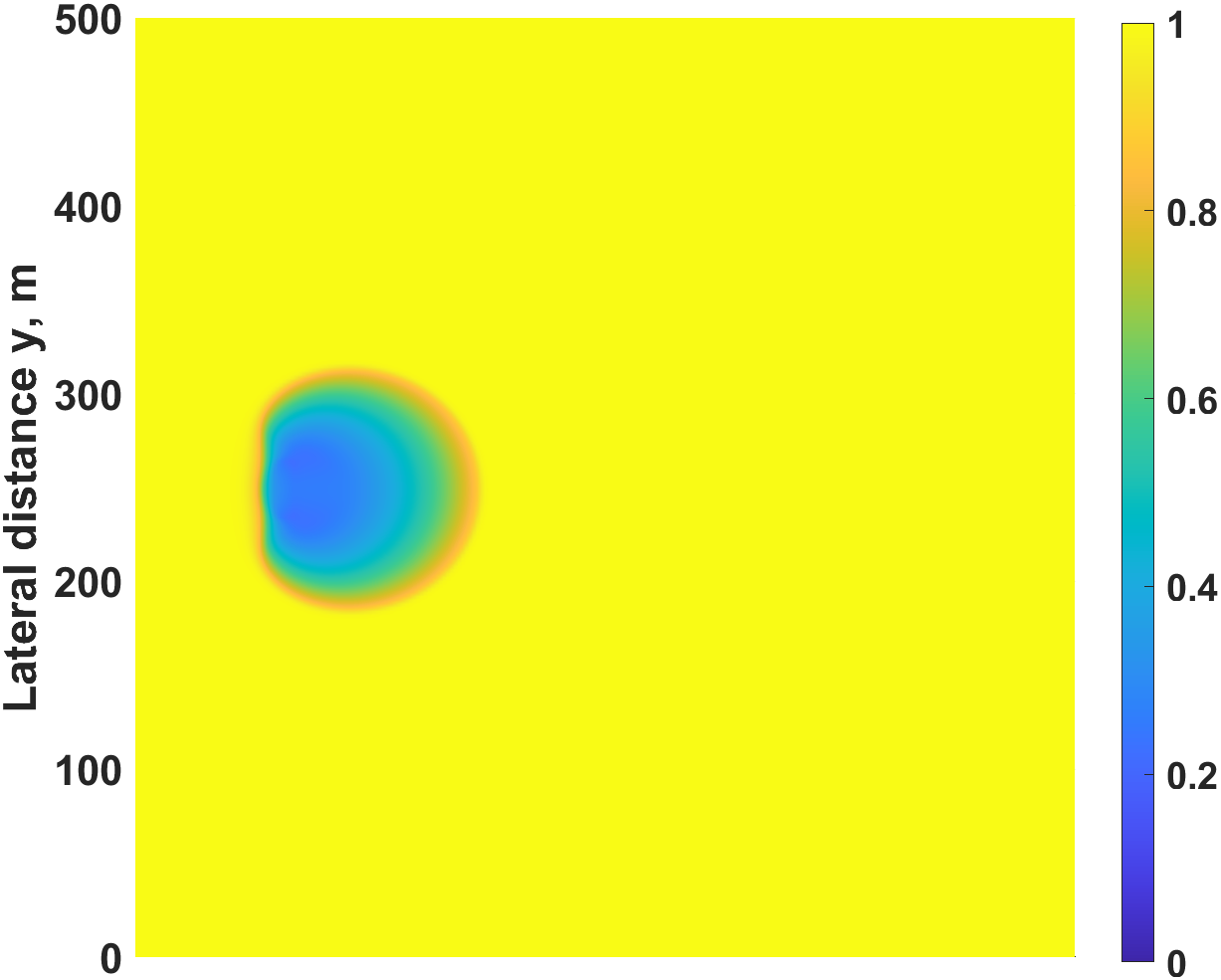} & &
\includegraphics[width=0.315\textwidth]{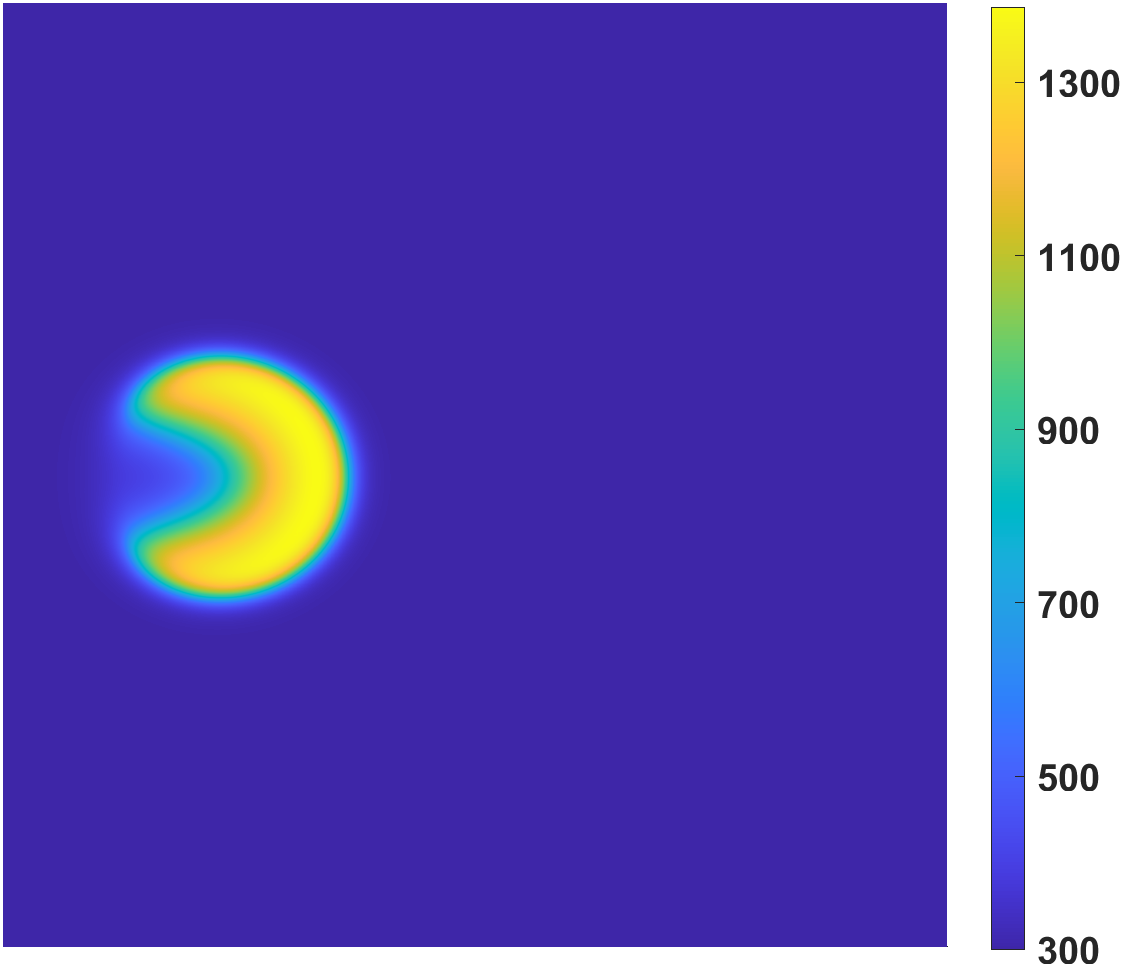} \\
\includegraphics[width=0.343\textwidth]{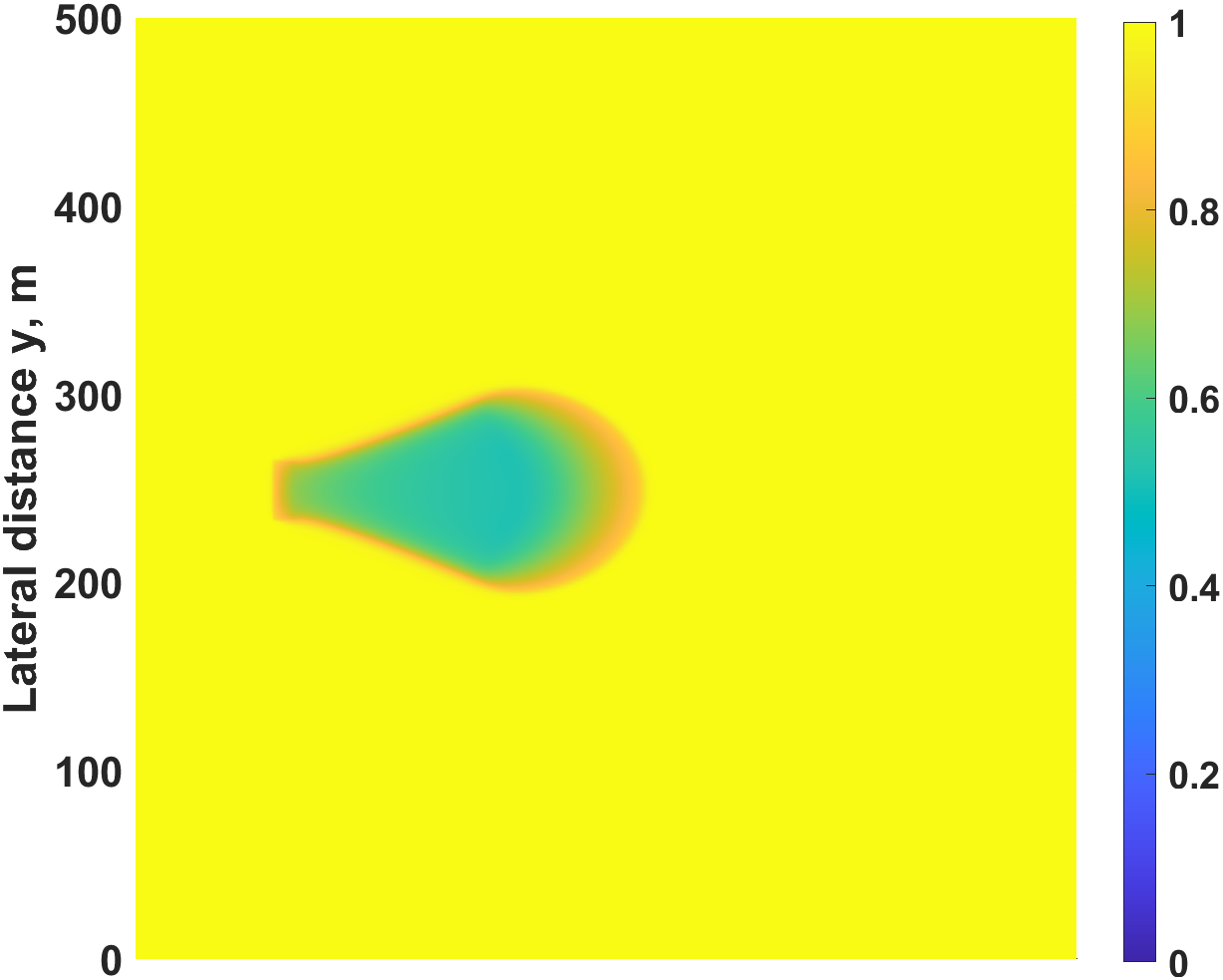} & &
\includegraphics[width=0.315\textwidth]{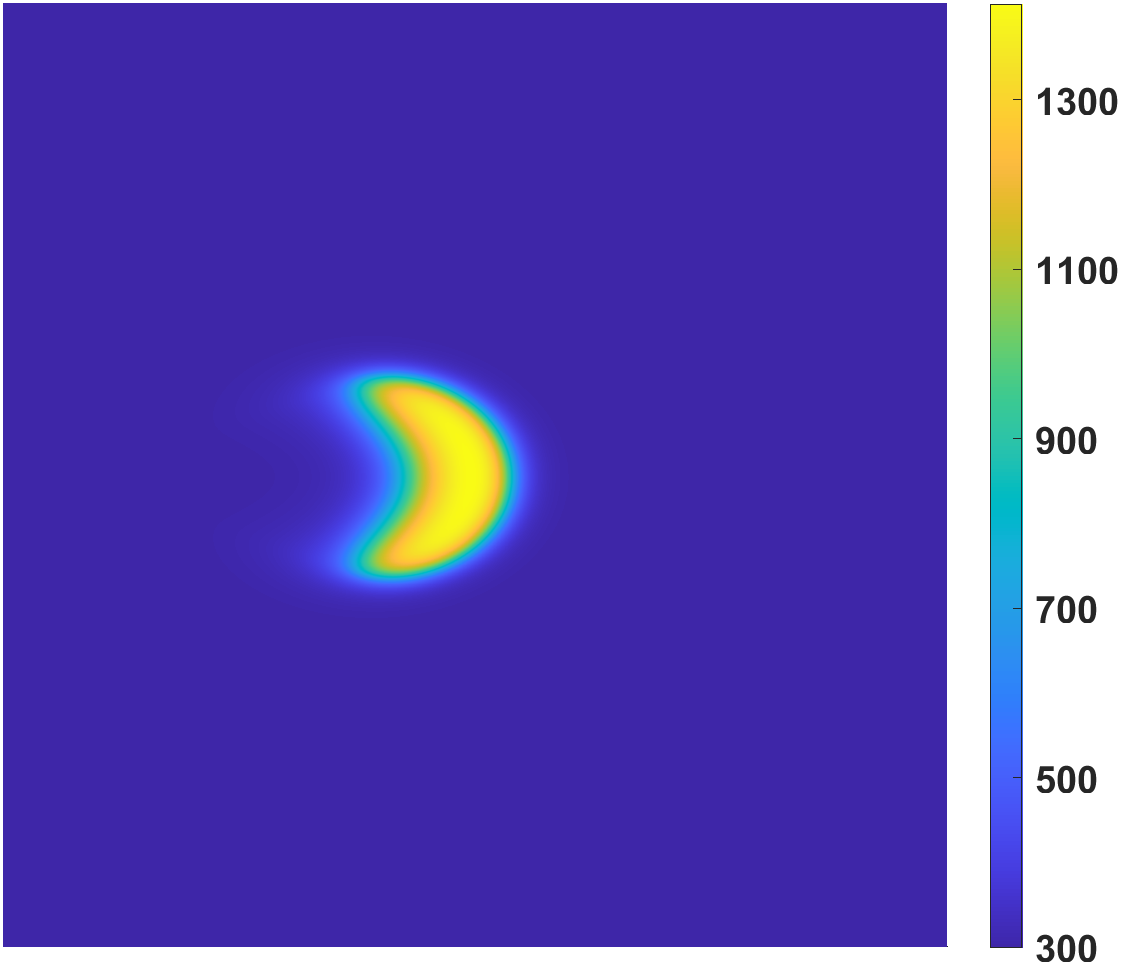} \\
\includegraphics[width=0.343\textwidth]{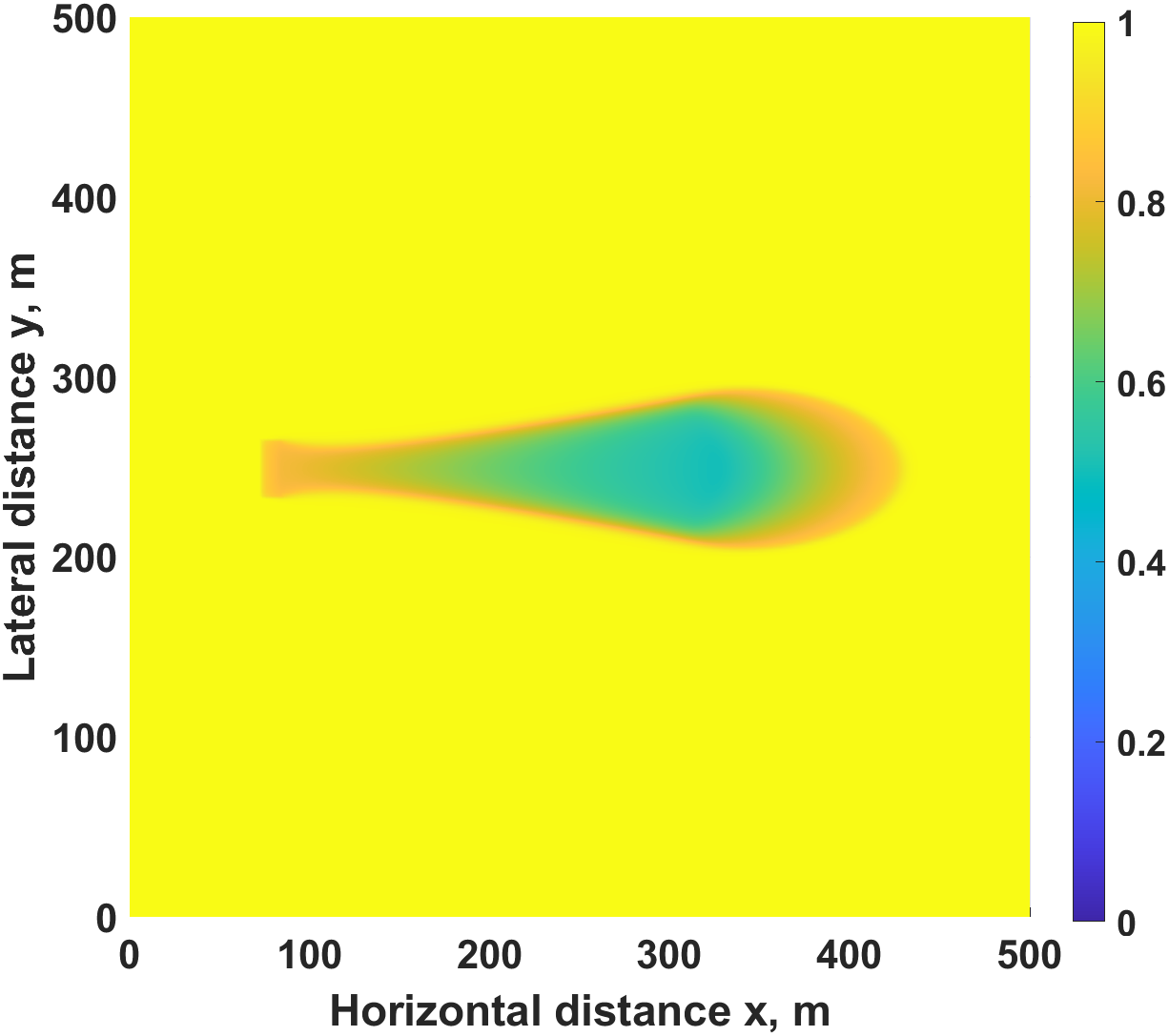} & &
\includegraphics[width=0.315\textwidth]{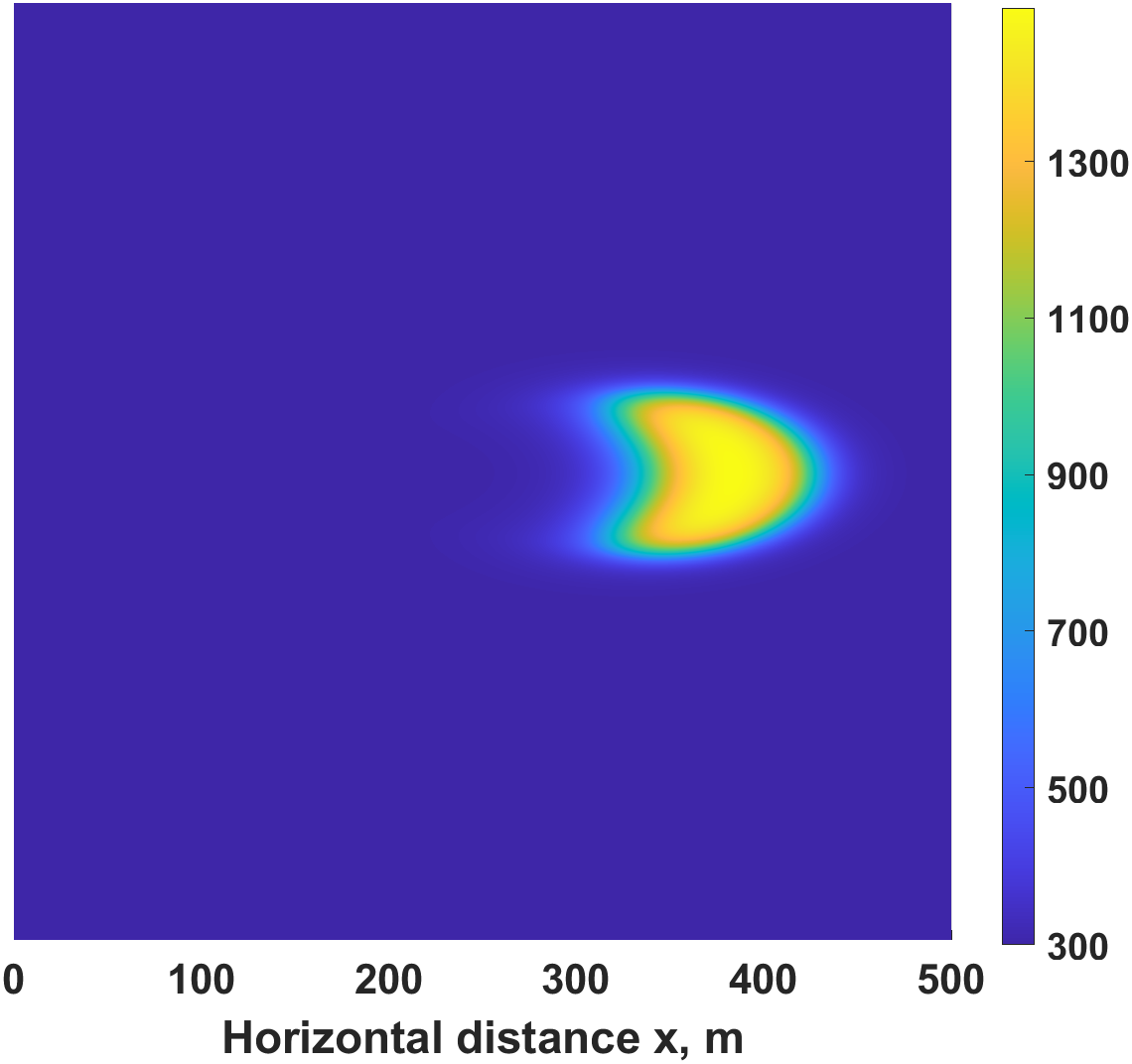} 
\end{tabular}
\caption{The state of the burning field, 900 s after ignition from a localized source with $x \in$ [75, 85] m and $y 
\in$ [235, 265] m. Left column: the remaining total mass fraction of solid. Right column: The spatial distribution of temperature. The rows are (from top to bottom) for $u_{10}=$ 0, 3, 6, and 10 m/s.}
\label{fig:FuelMassFraction}
\end{figure}

This and the following section present and discuss simulations of fire spread over a two-dimensional field. \Cref{fig:FuelMassFraction} shows the evolution of a fire, developing from a 10 m (width) by 30 m (length) ignition site after a lapse of $t=$ 900 s from ignition. Successive rows correspond to increasing wind velocity, whose direction is always at right angles to the initial fireline length. The left column shows the fraction of initial material remaining on the field, providing an overview of the fire-affected area. The right column displays the temperature distribution, highlighting the firefront's location and shape. The first row in \cref{fig:FuelMassFraction} corresponds to $u_{10}=$ 0 m/s, and, as expected, the fire-affected area is a perfect circle. As wind speed increases ($u_{10}=$ 3, 6, and 10 m/s in the 2nd, 3rd, and 4th rows, respectively), the shape of the fire-affected region becomes progressively narrower transversely and far more elongated in the wind direction \citep{AlexanderME1985}. An interesting feature of the left column of \cref{fig:FuelMassFraction} is the difference in fuel consumption. Thus, with no wind, maximum consumption occurs at the ignition site, and relatively little fuel remains there as a consequence of the slow burning process. This trend decreases with increasing wind speeds, and at higher velocities, maximum fuel consumption shifts with the advancing firefront, leaving the ignition site relatively rich in fuel.

\begin{figure}[tbp]
\centering \includegraphics[width=.4\textwidth]{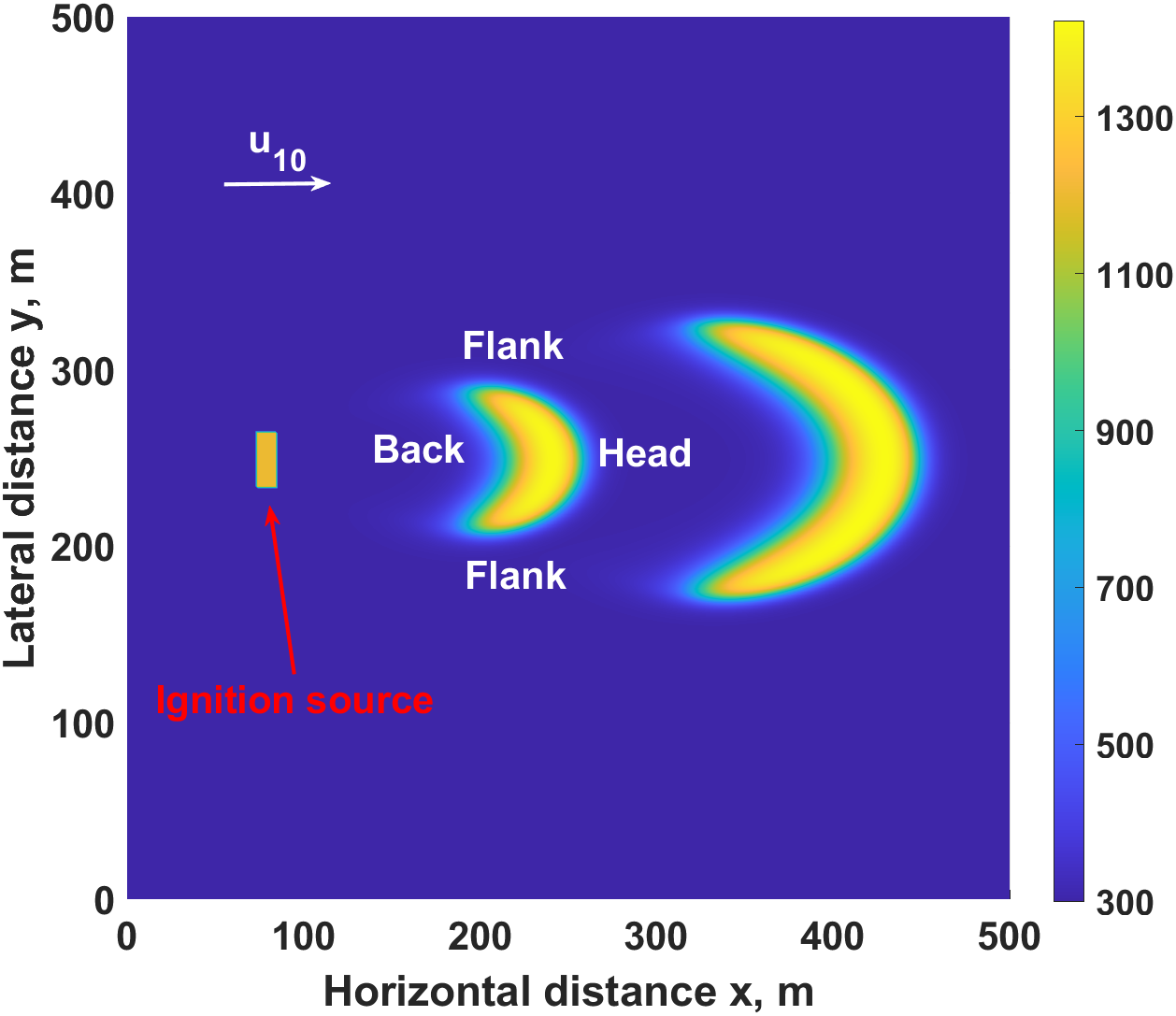}
\caption{Spatiotemporal evolution of the temperature profile at three time instants, $t=$ 0, 900, and 1800 s, for $u_{10}=$ 6 m/s.}
\label{fig:CombustionWave}
\end{figure}

The change in the firefront location and shape with increasing wind speed is depicted in the right column of \cref{fig:FuelMassFraction}. As expected, the firefront moves faster with stronger winds. Additionally, the shape transitions from symmetric to horseshoe and then to a parabola with increasing steepness. This tendency of the high-temperature zone to progress more rapidly in the direction of the wind compared to the transverse direction, leading to the parabolic shape of the firefront, is strongly supported by laboratory and field studies \citep{WallaceLFons1946,HalEAnderson1983,Clark2004}. The evolution of the firefront is more clearly observed in \cref{fig:CombustionWave}, which depicts the temperature profile for a wind speed of $u_{10}=$ 6 m/s at three time instants, $t=$ 0, 900, and 1800 s after ignition. The figure illustrates that the front and its flanks progress at distinctly different speeds, which--in combination with the rapid cooling behind the front due to the incoming cold wind--results in the pointed parabolic shape of the fireline. Furthermore, the semi-burned material left behind remains totally dry and will easily ignite if heated again.

Another set of simulations examines the effect of the initial fireline length on the {\rm (ROS)}. It has been observed that both field studies and simulations predict that the $\rm ROS$ is slower for a short fireline and increases asymptotically to a steady-state value as the fireline length increases \citep{CheneyNP1995, JMCanfield2014, FinneyMA2019}. Additionally, it is also recalled that the effect of fireline length has been included in the dispersion coefficient, \cref{eq:Dispersion}, in an attempt to model this behavior.

\begin{figure}[tbp]
\centering
\includegraphics[width=0.5\textwidth]{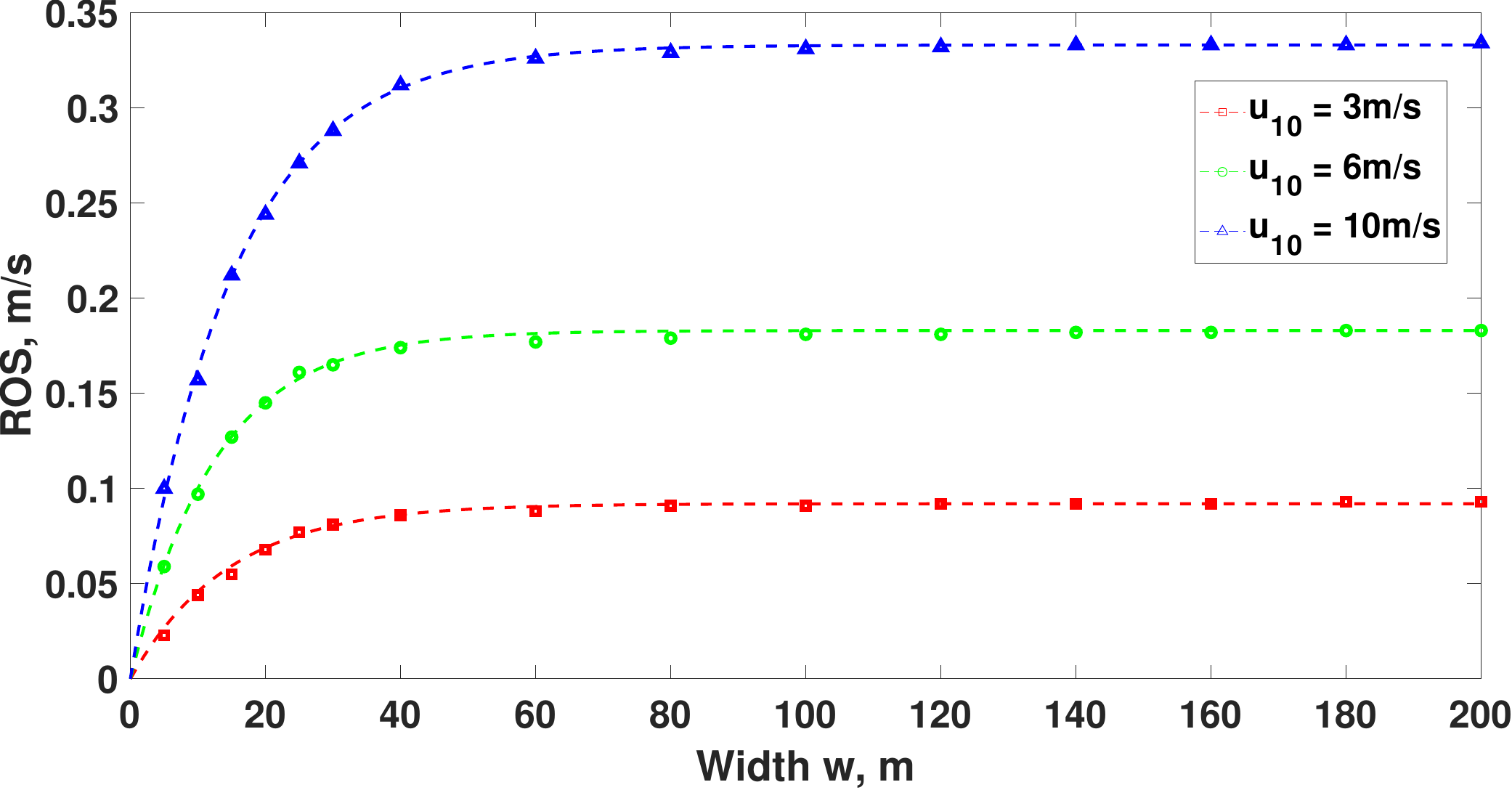}
\caption{The $\rm ROS$ as function of the width, $w$, of the fireline for wind speeds $u_{10}=$ 3, 6, and 10 m/s. Points are simulation results and lines the best-fit to \cref{eq:RosWidth}.}
\label{fig:ROS_width}
\end{figure}

The results of the existing model are presented in \cref{fig:ROS_width}. Three different wind speeds, $u_{10}=$ 3, 6, and 10 m/s are considered, and the initial length of the fireline in the transverse direction (normal to the wind) is varied within the interval $w \in$ [5, 200] m. It is evident from \cref{fig:ROS_width} that, for all cases, the $\rm ROS$ is accurately described by an equation of the form:
\begin{equation}
\label{eq:RosWidth}
{\rm ROS} = \lambda_1(1-e^{-\lambda_2 w}),
\end{equation}
with the best-fit values of the two coefficients given in \cref{table:ExponentialFunction}.
Coefficient $\lambda_1$ ([=] m/s) represents the asymptotic (quasi-steady-state) velocity of the combustion wave for a long fireline, ${\rm (ROS)}_{\infty}$. It is evidently determined by the strength of the wind, which remains the dominant influence \citep{WMell2007}. The coefficient $\lambda_2$ ([=] m$^{-1}$) influences the length, $w_0$, of the fireline beyond which the asymptotic value, ${\rm (ROS)}_{\infty}$, is practically reached. According to \cref{table:ExponentialFunction}, $\lambda_2$ depends very weakly on $u_{10}$, and a constant value $\lambda_2 \approx$ 7.3$\cdot$10$^{-2}$ m$^{-1}$ gives accurate results. By selecting a smaller value for parameter $\gamma_d$ in \cref{eq:Dispersion}, $a_2$ also decreases and the length $w_0$ increases.

The above estimate of the effect of limited fireline length compares very favorably with the predictions of the detailed 3D simulations of Pimont et al. \citep{Pimont2012}.These authors considered two fireline lengths, 20 and 50 m, respectively, and their computed rates of spread scale as ${\rm (ROS)}_{20m}=0.73-0.76\,{\rm (ROS)}_{50m}$ for $u_{10}=$ 5 and 12 m/s, respectively. The average ratio extracted from \cref{fig:ROS_width} for the wind speed range, $u_{10}=$ 3 - 10 m/s, is 0.78 $\pm$ 0.02.

\begin{table}[htbp]
\centering
\caption{Best-fit values for the parameters in \cref{eq:RosWidth}}
\label{table:ExponentialFunction}
\begin{tabular}{ccc}
\toprule
$u_{10}$ ([=] m/s) & $\lambda_1$ ([=] m/s) & $\lambda_2$ ([=] m$^{-1}$) \\
\midrule
3 & 0.09 & 6.91$\cdot10^{-2}$ \\
6 & 0.18 &7.93$\cdot10^{-2}$ \\
10 & 0.33 & 6.72$\cdot10^{-2}$ \\
\bottomrule
\end{tabular}
\end{table}

\subsection{Two-dimensional wildfire dynamics: fuel heterogeneity, firefront collision, and fuel breaks}
\label{sec:2D_Case_Studies}

\begin{figure}[tbp]
\centering 
\includegraphics[width=\textwidth]{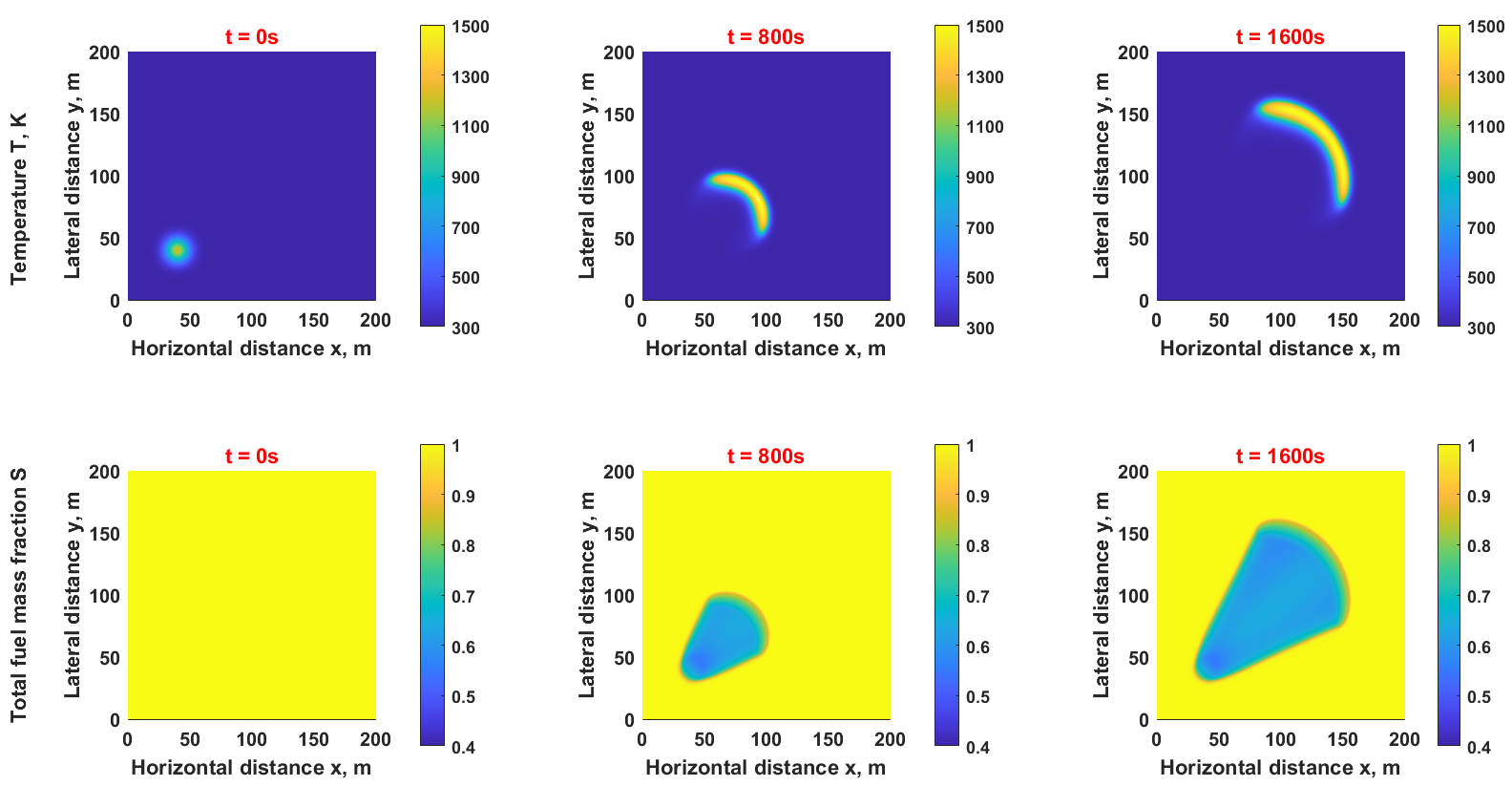}
\caption{Spatiotemporal evolution of temperature and total fuel due to uniform packing ratio, $\alpha =$ 0.002.}
\label{fig:Uniform_Packing_Ratio}
\end{figure}

The paper concludes with three case studies that represent preliminary efforts to simulate situations of great practical interest. They are all variations of the same base case, shown in \cref{fig:Uniform_Packing_Ratio}. It corresponds to wind speed $u_{10}=3\sqrt{2}$ m/s blowing in the direction of the diagonal ($u_{10,x}=u_{10,y}=$ 3 m/s) over a horizontal terrain of plantation with uniform packing ratio $\alpha=0.002$, dimensionless humidity $S_{1,0}=0.1$ and all other parameters as in \cref{table:ModelParameters}. Shown below are a few time instants, but the complete evolution scenarios are included as videos in the Supplementary Material.

\begin{figure}[tbp]
\centering 
\includegraphics[width=\textwidth]{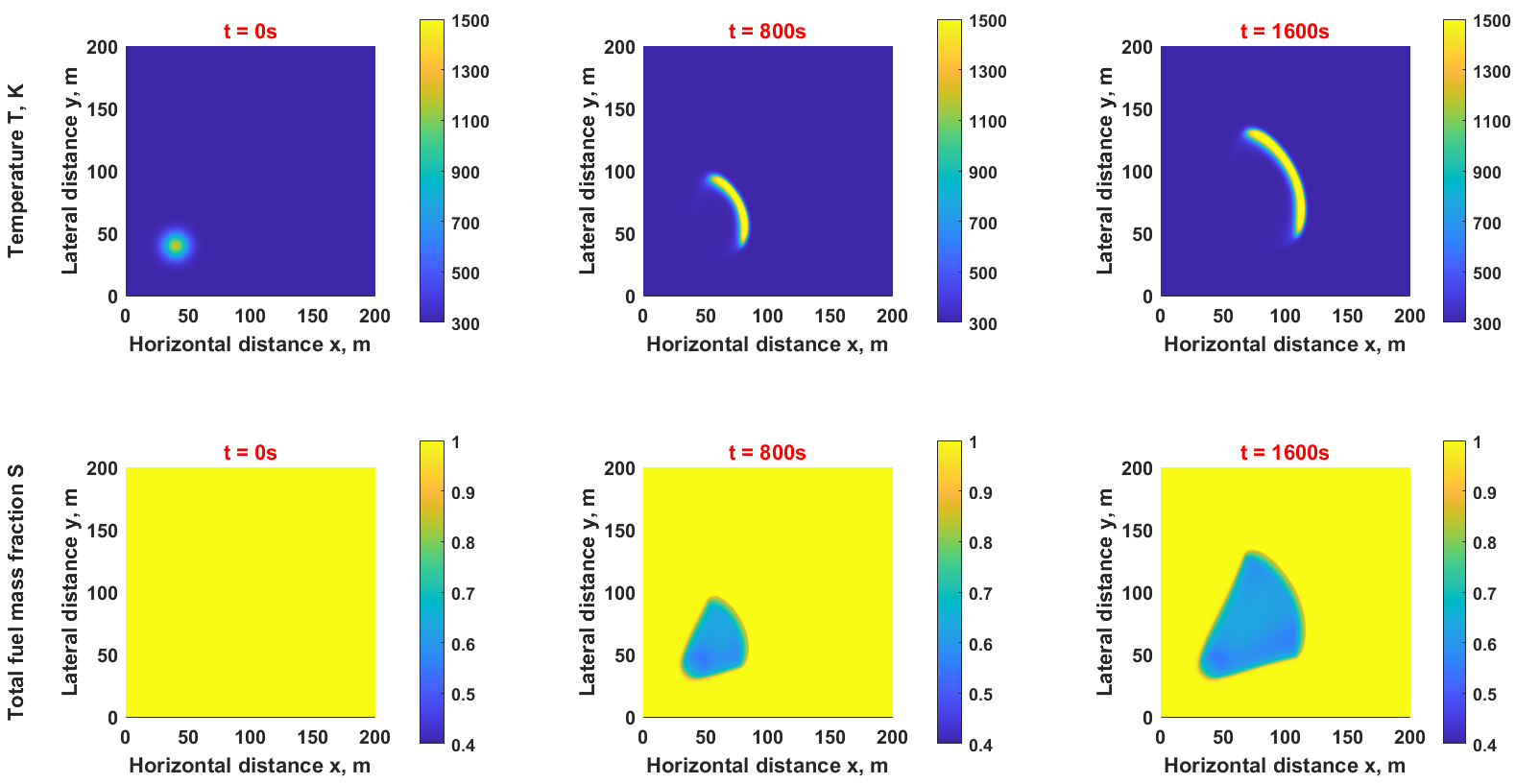}
\caption{Spatiotemporal evolution of temperature and total fuel due to variable packing ratio, $\alpha_1=$ 0.002 $\forall \, x<$ 60 m and $\alpha_2=$ 0.006 $\forall \, x \ge$ 60 m.}
\label{fig:Variable_Packing_Ratio}
\end{figure}

The first case study demonstrates the effect of spatial variation on packing ratio by considering the following distribution:  
\begin{equation}
\label{eq:Packing_ratio_var}
\alpha = \alpha(x) = \frac{\alpha_1 + \alpha_2}{2} + \frac{|\alpha_2 - \alpha_1|}{2}\tanh\left(\frac{x - x_{crit}}{10}\right),
\end{equation}
The values implemented are $\alpha_1=0.002$, and $\alpha_2=0.006$, representing two different zones in the $x-$direction with a smooth transition region centered at $x_{crit}=$ 60 m. \Cref{fig:Variable_Packing_Ratio} illustrates the time variation of fireline location (temperature distribution) and fire-affected region (remaining fuel distribution). Compared to the base case in \cref{fig:Uniform_Packing_Ratio}, the fire now progresses asymmetrically, expanding more intensely in the lateral ($y-$) direction, where the packing ratio remains small, and decelerating in the streamwise ($x-$) direction when entering the denser area. Though the fire-affected region is now small (due to the arrest in the rate of spread), fuel consumption is higher in the $x-$direction, as the denser area sustains combustion longer, leading locally to higher temperatures. (A variety of other interesting distributions can be envisioned--for example a denser patch surrounded by sparser plantations--and are left for future consideration.) 

\begin{figure}[tbp]
\centering 
\includegraphics[width=\textwidth]{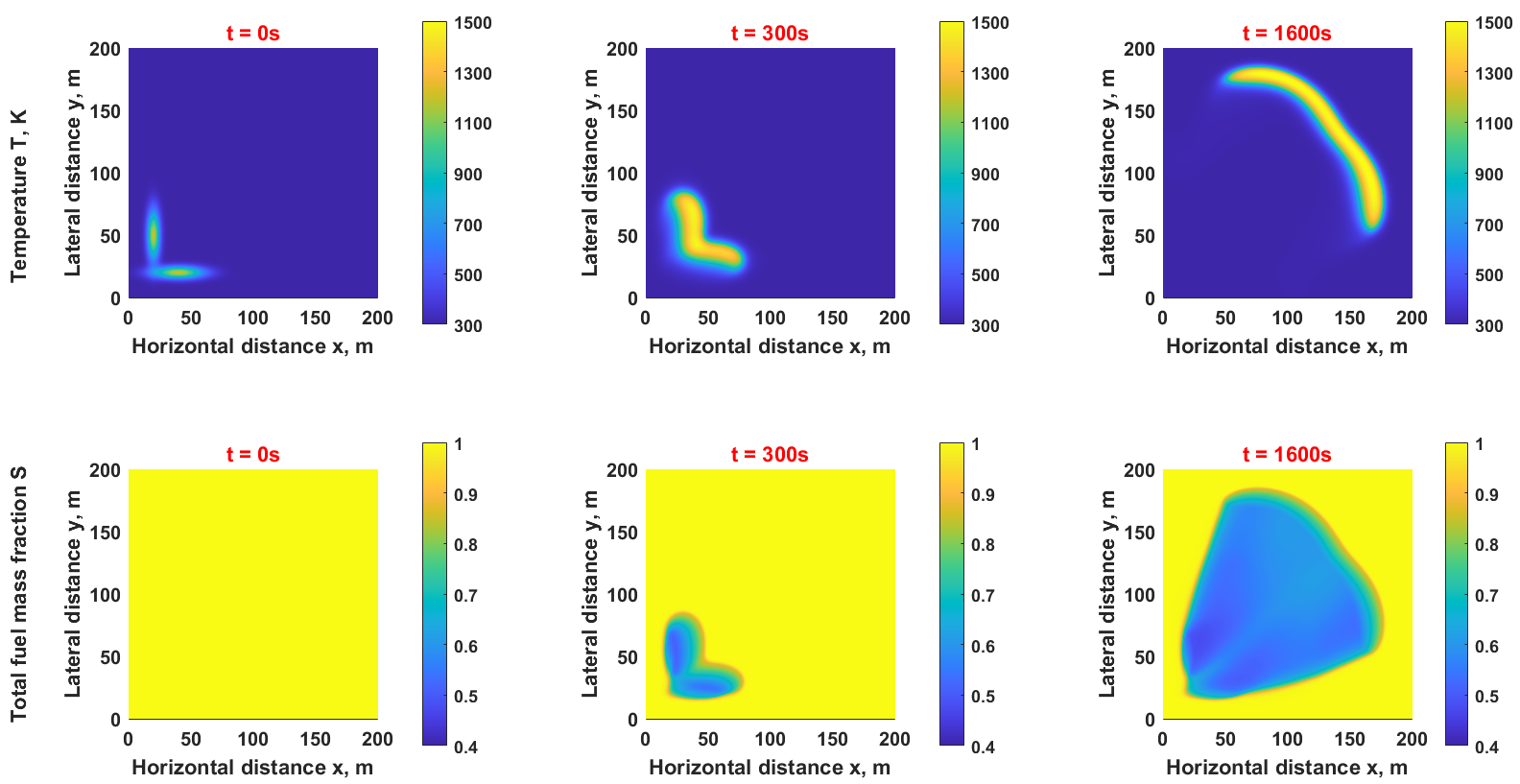}
\caption{Spatiotemporal evolution of temperature and total fuel due to the intersection of two firefronts.}
\label{fig:Firefront_Collision}
\end{figure}

One of the most important mechanisms enhancing wildfire spread is spot fire coalescence \citep{Koo2010}, which occurs when multiple fires burning in close proximity merge to form a single fireline. For example, when two straight firelines progressing in oblique directions intersect (L-shaped fires), the interaction significantly accelerates the rate of spread at the intersection zone \citep{Viegas2012}. This phenomenon is simulated in \cref{fig:Firefront_Collision} and the rapid progression of the merging zone to form the leading firefront is evident. Comparison with the base case, \cref{fig:Uniform_Packing_Ratio}, also indicates a drastically increased fire-affected region.

\begin{figure}[tbp]
\centering 
\includegraphics[width=\textwidth]{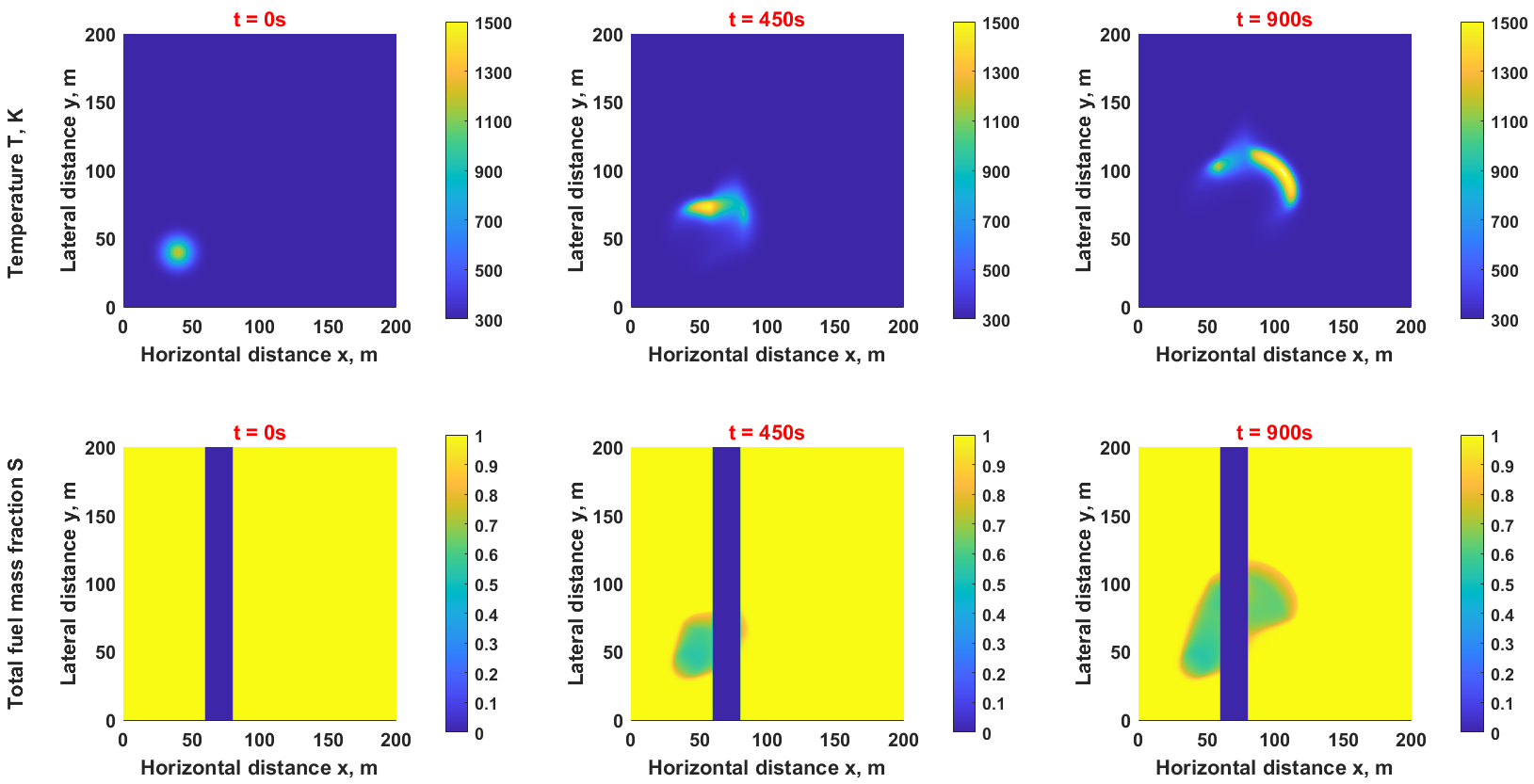}
\caption{Spatiotemporal evolution of temperature and total fuel due to a fuel break with $x \in$ [60, 80] m and $y \in$ [0, 200] m.}
\label{fig:Fuelbreak_1}
\end{figure}

\begin{figure}[tbp]
\centering 
\includegraphics[width=\textwidth]{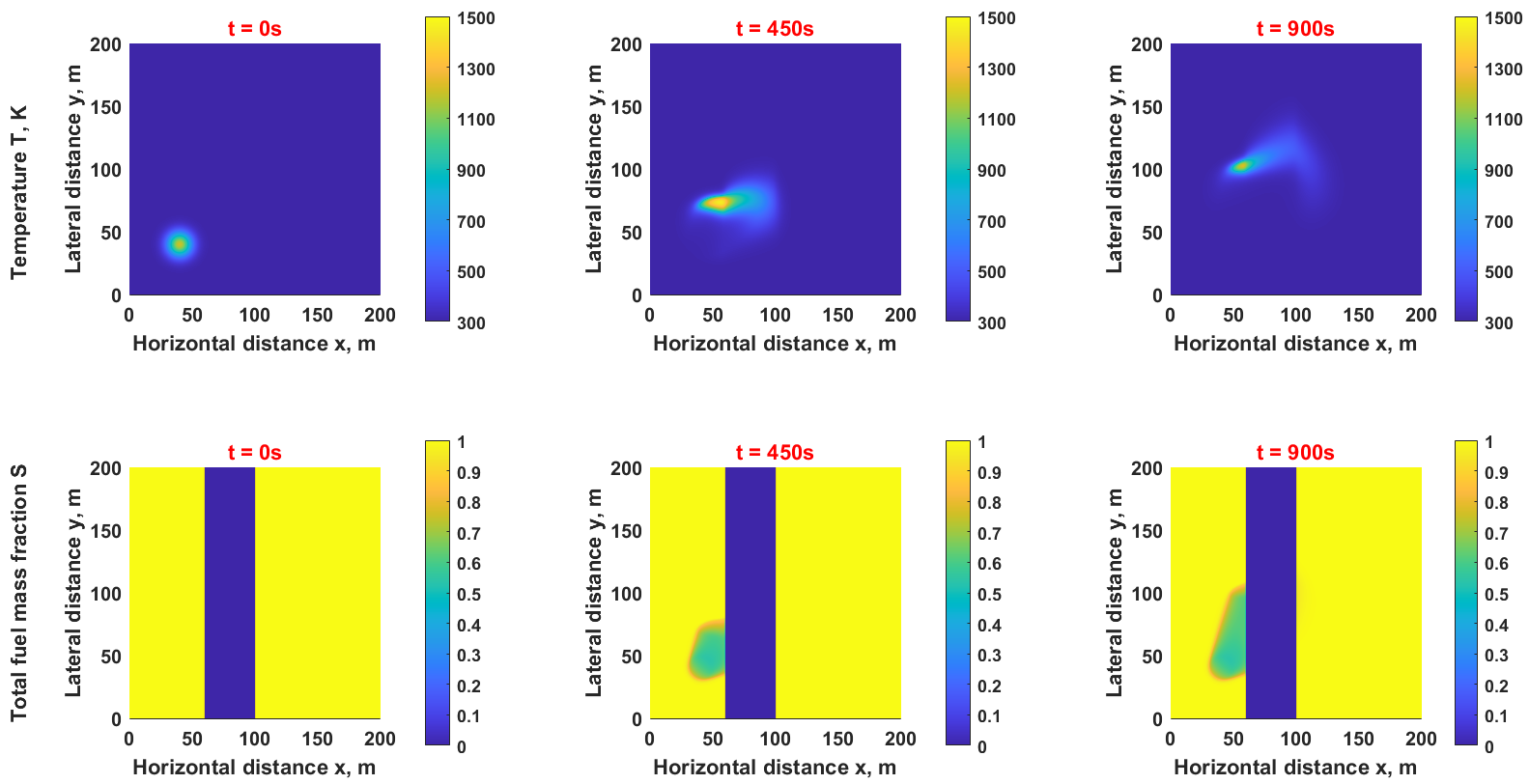}
\caption{Spatiotemporal evolution of temperature and total fuel due to a fuel break with $x \in$ [60, 100] m and $y \in$ [0, 200] m.}
\label{fig:Fuelbreak_2}
\end{figure}

The last case study considers the influence of fuel breaks, i.e. zones devoid of fuel. Fuel breaks serve to arrest fire expansion, or at least to provide a safer area for firefighters who combat the flames \citep{Agee2005} Thus, reliable estimation of the effect of their width is critical. \Cref{fig:Fuelbreak_1} and \cref{fig:Fuelbreak_2} show the predicted evolution of a firefront developing from a local ignition site, when it encounters a fire break of width 20 or 40 m respectively. In the first case, the fire propagates past the fuel break zone, while in the second it is arrested and extinguished.

\section{Conclusions and Outlook}
\label{sec:Conclusions}

A physics-based and interpretable model of fire propagation has been developed by invoking simplified reaction kinetics and energy conservation. The latter includes a convective contribution by the mean gaseous velocity through the canopy, a dispersion term that accounts for short-scale heat transfer by turbulence, buoyant currents, and radiation, and a term representing losses to the ambient by free convection and radiation. The mean velocity through the canopy is a key quantity for both the convection and dispersion terms. It is estimated from the effect of ambient wind, $u_{10}$, combining two extremes: (i) air flow through an intact canopy (wind-induced momentum dissipated as canopy drag) and (ii) air flow above totally burned ground (wind-induced momentum dissipated as rough wall drag).

In the present work, an off-line validation of the developed model is attempted by providing predictions for which benchmark data are available in the literature. Concerning fuel properties, it is shown that higher bulk density and fuel humidity lead to slower fire spread, in quantitative agreement with semi-empirical expressions in the literature. 
The effect of ambient wind is considered first for a straight firefront (1D propagation). With increasing wind speed, backward propagation of the firefront is decelerated and then extinguished, while propagation along the wind is intensified, varying roughly with the square of the speed, $u_{10}$. The combined effect of wind and terrain inclination is also considered, and predictions are found to be in satisfactory agreement with the results of the available detailed CFD simulations.

Fire propagation in two dimensions is considered next, starting with the evolution from a localized ignition site. With increasing wind speed, the firefront evolves from symmetric to horseshoe to parabolic, while the fire-affected region is increasingly elongated in the wind direction and restricted in the transverse. The effect of the length of the initial ignition front is also investigated, and the predicted reduction of the $\rm ROS$ for short firelines is found to be in quantitative agreement with the results of CFD attempts.

The investigation concludes with three case studies, which represent preliminary simulations of phenomena of great practical significance, fire propagation through heterogeneous plantation, oblique firefront collision and the interaction of a firefront with fuel break zones. In all these case studies, the predictions confirm qualitatively the expected behavior.

The developed model is envisioned as a component of a more general data-informed and quick-feedback simulation tool that will assist decisions in the management of wildfires. To this end, its present capabilities need to be improved in at least two directions. First, the simulated effect of terrain inclination needs to be generalized for arbitrary topography. Second, the flammable material needs to be diversified in accordance with the available standard fuel models \citep{Anderson1982, Scott2005}. These models have been developed for application with Rothermel's surface fire spread correlation, so they produce fuel properties as averages of those of the constituent species \citep{Andrews2018}. One potential advantage of the present methodology is that the consumption of each constituent may be independently followed (by appropriate combustion kinetics), leading to a more realistic fire evolution scenario.


%\section{}\label{}


\bibliographystyle{elsarticle-num}

\bibliography{Wildfire_modeling}

%% Biography
%% Here goes the biography details.

%% Here goes the biography details.

\end{document}